\documentclass[%
twocolumn,
superscriptaddress,
 amsmath,amssymb,
pra,
floatfix,
]{revtex4-2}

\usepackage[english]{babel}
\usepackage[utf8]{inputenc}
\usepackage{dcolumn}
\usepackage{bm}
\usepackage{amsthm}
\usepackage{amssymb}
\usepackage{mathtools}
\usepackage{physics}
\usepackage{xcolor}
\usepackage{graphicx}
\usepackage{float}
\usepackage{adjustbox}
\usepackage{placeins}
\usepackage[T1]{fontenc}
\usepackage{lipsum}
\usepackage{csquotes}
\usepackage{comment}
\usepackage{xr}
\usepackage{makecell}
\usepackage{ mathrsfs }
\usepackage{url}
\usepackage{capt-of}

\newcommand{\mxy}{\ensuremath{\vec{m}_{xy}}} 
 
\newcommand{\feo}{\ensuremath{\alpha}-Fe\ensuremath{_2}O\ensuremath{_3}}
\newcommand{\dmxy}{\ensuremath{\vec{\nabla} \cdot \vec{m}_{xy}}}

\newcommand{\nm}{\ensuremath{\mathrm{nm}}}
\makeatletter
\newcommand*{\addFileDependency}[1]{
  \typeout{(#1)}
  \@addtofilelist{#1}
  \IfFileExists{#1}{}{\typeout{No file #1.}}
}
\makeatother

\begin{document}
\title{\Large Revealing Emergent Magnetic Charge in an Antiferromagnet with Diamond Quantum Magnetometry} 

\author{Anthony K. C. Tan}
\thanks{Contributed equally to this work}
\email{a.tan@imperial.ac.uk}
\affiliation{Cavendish Laboratory, University of Cambridge, J. J. Thomson Avenue, Cambridge, CB3 0HE, UK}

\author{Hariom Jani}
\thanks{Contributed equally to this work}
\email{hariom.jani@physics.ox.ac.uk}
\affiliation{Department of Physics, National University of Singapore, Singapore, Singapore}
\affiliation{Clarendon Laboratory, Department of Physics, University of Oxford, Oxford, UK}

\author{Michael H{\"o}gen}
\thanks{Contributed equally to this work}
\affiliation{Cavendish Laboratory, University of Cambridge, J. J. Thomson Avenue, Cambridge, CB3 0HE, UK}

\author{Lucio Stefan}
\affiliation{Center for Hybrid Quantum Networks (Hy-Q), Niels Bohr Institute, University of Copenhagen, Copenhagen, Denmark}

\author{Claudio Castelnovo}
\affiliation{Cavendish Laboratory, University of Cambridge, J. J. Thomson Avenue, Cambridge, CB3 0HE, UK}

\author{Daniel Braund}
\affiliation{Cavendish Laboratory, University of Cambridge, J. J. Thomson Avenue, Cambridge, CB3 0HE, UK}

\author{Alexandra Geim}
\affiliation{Cavendish Laboratory, University of Cambridge, J. J. Thomson Avenue, Cambridge, CB3 0HE, UK}

\author{Matthew S. G. Feuer}
\affiliation{Cavendish Laboratory, University of Cambridge, J. J. Thomson Avenue, Cambridge, CB3 0HE, UK}

\author{Helena S. Knowles}
\affiliation{Cavendish Laboratory, University of Cambridge, J. J. Thomson Avenue, Cambridge, CB3 0HE, UK}

\author{Ariando Ariando}
\affiliation{Department of Physics, National University of Singapore, Singapore, Singapore}

\author{Paolo G. Radaelli}
\email{paolo.radaelli@physics.ox.ac.uk}
\affiliation{Clarendon Laboratory, Department of Physics, University of Oxford, Oxford, UK}

\author{Mete Atat{\"u}re}
\email{ma424@cam.ac.uk}
\affiliation{Cavendish Laboratory, University of Cambridge, J. J. Thomson Avenue, Cambridge, CB3 0HE, UK}


\begin{abstract}

Whirling topological textures play a key role in exotic phases of magnetic materials and offer promise for logic and memory applications. In antiferromagnets, these textures exhibit enhanced stability and faster dynamics with respect to ferromagnetic counterparts, but they are also difficult to study due to their vanishing net magnetic moment. One technique that meets the demand of highly sensitive vectorial magnetic field sensing with negligible backaction is diamond quantum magnetometry. Here, we show that the archetypal antiferromagnet, hematite, hosts a rich tapestry of monopolar, dipolar and quadrupolar emergent magnetic charge distributions. The direct readout of the previously inaccessible vorticity of an antiferromagnetic spin texture provides the crucial connection to its magnetic charge through a duality relation. Our work defines a novel paradigmatic class of magnetic systems to explore two-dimensional monopolar physics, and highlights the transformative role that diamond quantum magnetometry could play in exploring emergent phenomena in quantum materials.
\end{abstract}
\maketitle

\section{Introduction}
Topologically protected states in magnetic materials are promising candidates for novel spintronics architectures, such as neuromorphic computing \cite{Prychynenko2017,bourianoff2018potential}. In particular, topological textures in antiferromagnets (AFMs) could provide additional advantages over their ferromagnetic counterparts including enhanced stability, as well as faster and richer dynamics \cite{Keffer1952,Ross2015,Wang2019,Zhang2016,Barker2016,Velkov2016, Gobel2017,Buhl2017,Shiino2016,Gomonay2016,Hals2011,Selzer2016,psaroudaki2021skyrmion,lebrun2018tunable}. The vanishing net moment brings forth its own challenges, particularly making the detection of AFM textures difficult. Synchrotron-based dichroic X-ray techniques, sensitive to the staggered magnetization, are at the imaging forefront and have for the first time revealed the existence of two-dimensional topological AFM spin textures in hematite, \feo~\cite{Jani2021,Chmiel2018}. This powerful technique, however, comes with the caveat that it is insensitive to the sign of the staggered magnetization and thus the associated vorticity, i.e. the actual whirling of the spin textures, is not observed. 

When viewed instead through the lens of the canted magnetization, we uncover weak but finite magnetic fields emanating from the \emph{twists} of the staggered moments. The presence of such magnetic fields can equivalently be described by the magnetic analogue of Gauss’s law \cite{jackson1975classical}, therefore pointing to the existence of a distribution of emergent magnetic charges in a topologically-rich AFM landscape. The state of the art vectorial weak-field sensing of diamond quantum magnetometry (DQM), employing a single nitrogen-vacancy (NV) colour centre as a point field sensor \cite{Gross2017,Chauleau2020a,Finco2022,Wornle2021,Hedrich2021}, therefore puts it in a unique position to study the above proposed conceptual idea of emergent magnetic charges in a completely new class of magnetic materials -- canted AFMs.

In this letter, we demonstrate DQM imaging of topological textures in the archetypal AFM \feo~ and we show that these AFM textures host a rich tapestry of magnetic charge distribution.  In particular, the duality relation between staggered vorticity and magnetic charge allows us to associate the AFM Bloch meron with a spatially extended \emph{emergent magnetic monopole}. Distinct from emergent magnetic monopoles in other realizations, such as spin ice\cite{Castelnovo2008}, we observe here that the positively and negatively charged monopolar textures are topologically equivalent, while their topological antiparticle has a magnetic quadrupolar character.  In addition to being a promising tool to investigate topological AFM textures, our results demonstrate the DQM potential for discovering emergent phenomena in model magnets. 

\begin{figure*}
    \centering
    \includegraphics[width =  0.9\textwidth]{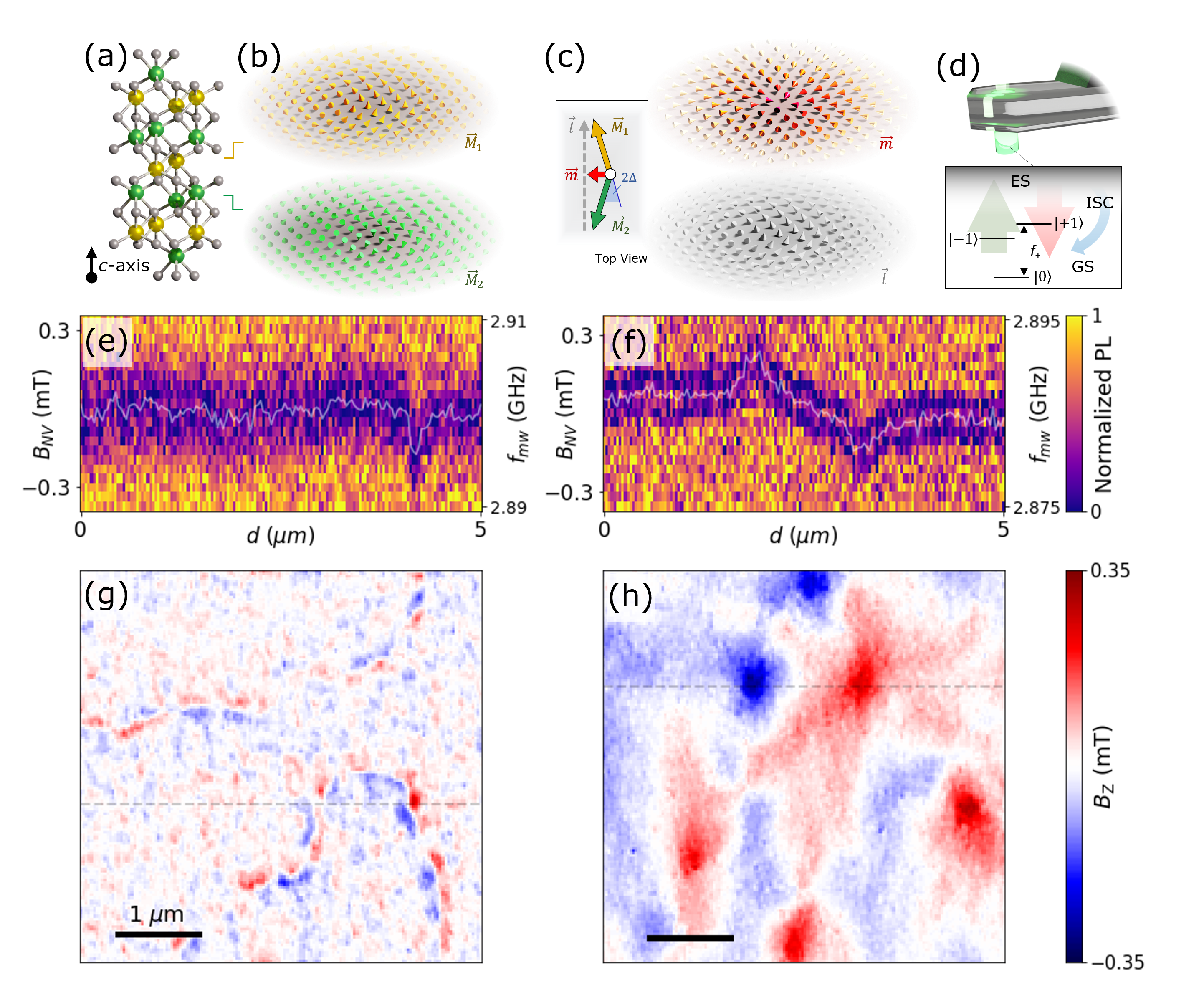}
    \caption{ \textbf{Signatures of emergent magnetic field  in hematite \feo.} \textbf{(a)} Atomic structure of \feo~(Fe and O atoms in yellow/green and grey spheres respectively). \textbf{(b)} Discrete representation of the alternating FM sublattice magnetisation $\vec{M_1}$ (yellow cones) and $\vec{M_2}$ (green cones) with AFM coupling along the $c$-axis shown in (a). \textbf{(c)} Illustration of the whirling staggered magnetisation, $\vec{l}$ (grey cones), forming an anti-clockwise a-Bloch meron, and the resultant canted magnetic moment, $\vec{m}$ (red cones). The insert shows the relationship between $\vec{l}$, $\vec{m}$, $\vec{M_1}$, $\vec{M_2}$ and the canting angle $\Delta$. \textbf{(d)} A scanning diamond sensor with a single NV centre maps out the $B$-field generated near the sample surface. The insert shows the energy diagram of the NV ground state (GS) $\ket{\pm 1}$ and $\ket{0}$ sublevels. A microwave field drives the GS spin transition while a 532-nm laser excites the NV to the excited state (ES) (green arrow). The NV then undergoes a radiative decay to GS (red arrow) or a non-radiative and spin selective path via the intersystem crossing (ISC) (blue arrow) enabling ODMR acquisition. \textbf{(e, f)} ODMR along the fast scan direction, measured on \feo~ thin film at $T=4~\mathrm{K}$ (e) and across $T_{\mathrm{M}}$ at $300~\mathrm{K}$ (f). The fitted $f_+$ ($B_{\mathrm{NV}})$ is plotted as a white line in each panel. \textbf{(g, h)} $B_z$ images retrieved from fitted $B_{\mathrm{NV}}$ maps reveal distinct field signatures across $T_{\mathrm{M}}$. Dashed lines in (g) and (h) correspond to fitted $B_{\mathrm{NV}}$ traces in (e) and (f), respectively. Scale bars are $1~\mu \mathrm{m}$.}
    \label{fig:intro}
\end{figure*}
\section{Details of \feo~and DQM}
Hematite \feo~is an AFM oxide insulator, which hosts a wide family of topological spin textures \cite{Jani2021,oxide_skyrmions,Fe2O3_dmi}. Figure \ref{fig:intro}(a) illustrates the trigonal corundum atomic structure of \feo. It comprises a stack of anti-parallel FM sublattices along the $c$-axis, with magnetisation textures $\vec{M_1}$ (yellow) and $\vec{M_2}$ (green), as shown in Fig. \ref{fig:intro}(b).  Spin re-orientation occurs at the Morin transition temperature, $T_\mathrm{M}\sim 200~\mathrm{K}$ \cite{Jani2021}(see supplementary note S1); below and above $T_\mathrm{M}$, the FM sublattices lie predominantly out-of-plane and in-plane, respectively. The N\'{e}el vector $\vec{l}~=\vec{M_1}-\vec{M_2}$ characterises the AFM order, while $\vec{m}~= \vec{M_1}+\vec{M_2}$ is the net magnetisation, as shown in Fig. \ref{fig:intro}(c).  Above $T_\mathrm{M}$, $\vec{m}$ has a predominantly in-plane orientation with an average magnitude $m_\Delta\sim2\times10^3~\mathrm{A/m}$. This weak magnetisation is due to the slight in-plane canting of angle $\Delta ~$ $\sim1.1~\mathrm{mrad}$ (see S1) between $\vec{M_1}$ and $\vec{M_2}$, as a consequence of the intrinsic bulk Dzyalonshinskii–Moriya interaction (DMI) vector along the $c$-axis\cite{Thoma2021}(see S5). Consequently, $\vec{m}$ lies in-plane and satisfies the relation $\vec{m} \cdot \vec{l}=0$. Since $m_\Delta$ is around three orders of magnitude weaker than $|\vec{l}|$, this canting-induced weak magnetisation has no discernible effect on the AFM character of \feo. Finally, $\vec{m}$ vanishes in situations where $\vec{l}$ turns out-of-plane below $T_\mathrm{M}$ or due to the local formation of AFM spin textures.  

We quantify the magnetic-field distribution from these spin textures via a custom-built cryogenic DQM (see S2). Figure \ref{fig:intro}(d) illustrates our diamond cantilever probe hosting a single NV centre, which is scanned at a constant height above the sample surface. The NV centre is an atomistic single spin defect with a paramagnetic ground-state manifold and state-selective optical transitions. This allows the Zeeman splitting between the ground states $\ket{\pm1}$ and $\ket{0}$ to be probed with a microwave frequency $f_\mathrm{mw}$ sweep and optical excitation, i.e. via optically detected magnetic resonance (ODMR). Consequently, in the weak-field approximation\cite{Rondin2014} with negligible strain, we infer the local $B$-field projected onto the NV-axis ($B_{\mathrm{NV}}$) from the energy difference between $\ket{0}$ and $\ket{+1}$ given by $\Delta E_+= h( |f_{+}-D|- \tilde{\gamma} B_\mathrm{bias}) = h \tilde{\gamma} B_{\mathrm{NV}}$, where $h$ is Planck's constant, $f_{+}$ is the resonant frequency corresponding to the transition, $D\approx2.87~\mathrm{GHz}$ and $\tilde{\gamma} = 28~\mathrm{MHz/mT}$. A small bias field $B_\mathrm{bias}$ $\approx 0.5~\mathrm{mT}$ is applied along the NV-axis to enable extraction of the field orientation. Figures \ref{fig:intro}(e) and (f) illustrate the typical variation in ODMR frequency across a linescan over the \feo~surface at $T= 4~\mathrm{K}$ and $300~\mathrm{K}$, respectively. The colour plot displays the amplitude of the ODMR signal, while the white curves demarcate $f_+$ used to extract $B_\mathrm{NV}$. 
An ODMR raster scan across the sample surface provides a $B_\mathrm{NV}$ image. We then transform to the laboratory coordinates ($B_{x,y,z}$) via the Fourier reconstruction technique\cite{Lima2009}, where $z$ coincides with the $c$-axis of the crystal. Figures \ref{fig:intro}(g) and (h) present example images of $B_z$ collected at $4~\mathrm{K}$ and $300~\mathrm{K}$, respectively. The qualitative differences observed in these images already reveal distinct magnetic phases for temperatures below and above $T_{\mathrm{M}}$ \cite{Jani2021}. The $B_z$ image below $T_\mathrm{M}$ is generally weak, consistent with the absence of net magnetisation. By contrast, the $B_z$ image above $T_\mathrm{M}$ shows strong and distinct features, consistent with the expectations of non-zero net magnetisation in this phase.

\section*{Emergent Properties in \feo}
To gain a physical interpretation of the underlying magnetisation distribution from the measured $B_z$ images, we begin from a thin-film approximation given by (see S4.1):
\begin{equation}
    B_z = \alpha_{xy}(t,d) * \dmxy  + \alpha_{z}(t,d) * \nabla ^2 m_z~,
    \label{eq:bz_realspace}
\end{equation}
where $*$ indicates convolution, $\alpha_{i}$ ($i = xy, z$) are effective point spread functions\cite{Dovzhenko2018}, $t$ is the film thickness, $d$ is the height above the film surface, and $\mxy$ and $m_z$ are the in-plane and out-of-plane components of $\vec{m}$. $\mxy$ and $m_z$ contribute to $B_z$ through the divergence and Laplacian, respectively. The $\alpha_{i}$ account for the $B$-field decay above the surface, acting as blurring kernels with their size of order $d$. Hence, the spatial resolution of our DQM imaging technique is set by the effective NV-sample distance,  $d_\mathrm{NV}$ (see S3). Due to the symmetry of the DMI in \feo (see S5), $m_z=0$ and $\mxy\neq 0$, rendering the second term in Eq. \ref{eq:bz_realspace} identically zero. Therefore, $B_z$ images are the \emph{divergence of the canted magnetisation} $\dmxy$, convolved with $\alpha_{xy}$.  Moreover, for a $\hat{z}$-oriented DMI, the net magnetisation is also given by $\vec{m}=\Delta (\hat{z}\times \vec{l})$, where $\Delta$ is the DMI-set canting angle. This yields an additional expression, namely $\dmxy = \Delta [\hat{z} \cdot (\vec{\nabla} \cross \vec{l}~)]$ (see S5).  The striking consequence of this expression is that $B_z$ images also offer a projected measure of the \emph{staggered vorticity}, i.e. the \emph{curl of the N\'{e}el vector}, $\vec{\mathcal{V}} =\vec{\nabla} \cross \vec{l}$.
 \begin{figure*}
    \centering
    \includegraphics[width =  1\textwidth]{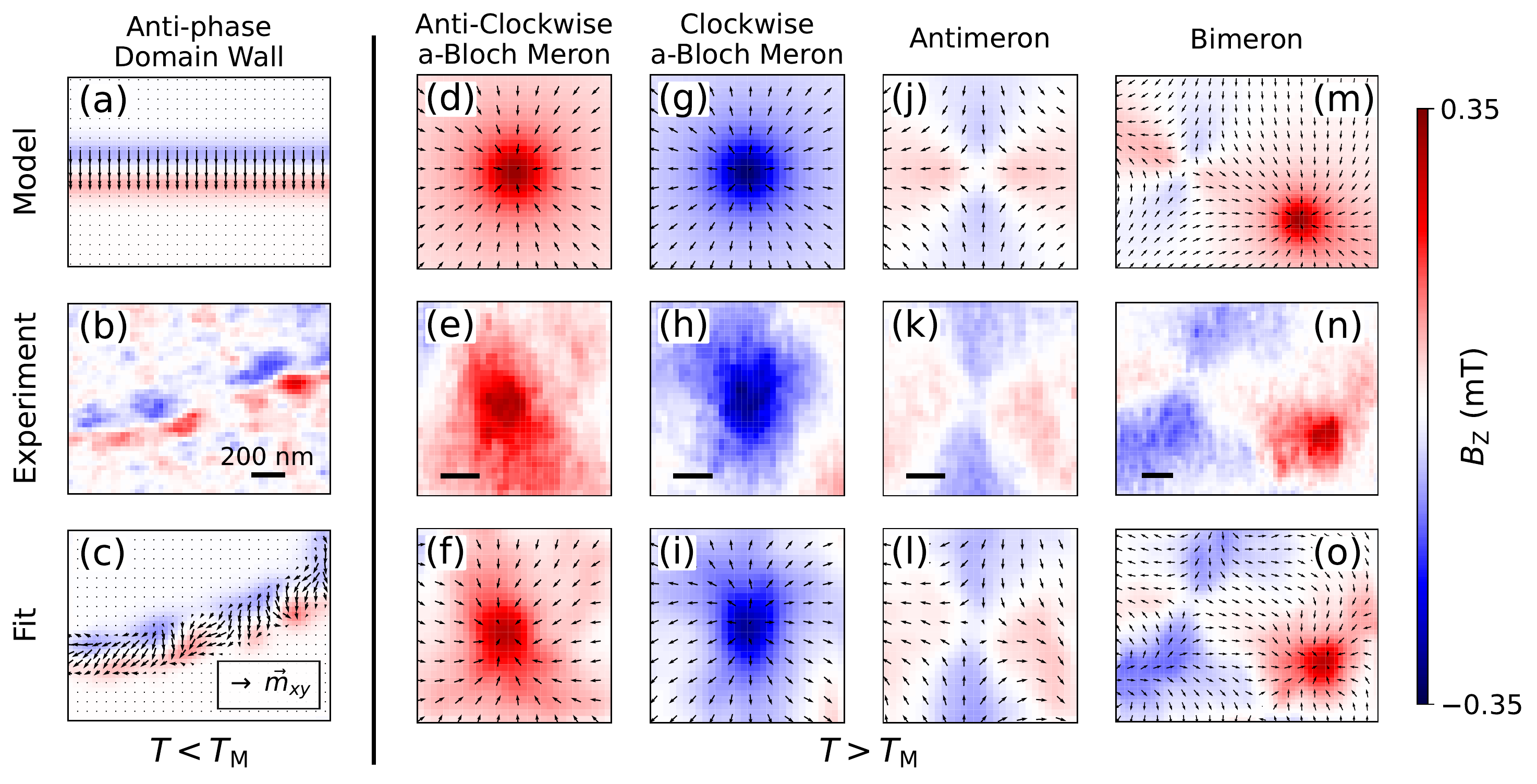}
    \caption{ \textbf{Classification of topological AFM textures via DQM.} \textbf{(a-c)} Topological AFM textures observed below $T_{\mathrm{M}}$. Distinct $B_z$ signature of an ADW simulated (a) and measured (b) above the sample surface. The reconstructed \mxy~ (black arrows) from (b) and its $B_z$ distribution are given in (c). \textbf{(d-o)} Topological AFM textures observed above $T_{\mathrm{M}}$. Simulated (d) and measured (e) $B_z$ field signatures of an anti-clockwise a-Bloch meron and (f) the reconstructed \mxy~(black arrows) of (e) and its associated $B_z$. Similarly, the simulated and measured $B_z$ signatures, and the \mxy~reconstruction of a clockwise a-Bloch meron, an anti-meron and a bimeron is given in panels (g-i), (j-l), (m-o) respectively. Details of \mxy~reconstruction are given in Ref. (see S7). Scale bars are $200~\mathrm{nm}$.}
    \label{fig:field}
\end{figure*}
\section*{Characteristic Field Signatures and Vorticity Readout}
With the gained insight on the relationship between the $B_z$ images and \mxy, we show next the $B_z$ images obtained in Fig. \ref{fig:intro} are produced by AFM antiphase domain walls (ADWs), merons, antimerons and bimerons -- consistent with their recent observation in \feo \cite{Jani2021}. Below $T_\mathrm{M}$, we model the $B_z$ images with a linear AFM domain wall ansatz \cite{Jani2021}, characterised by width $w$ and phase  $\xi_a$ (see S5.1). The phase $\xi_a$ controls the spatial variation of $\vec{l}$, resulting in an AFM N\'eel (a-N\'eel) or an AFM Bloch (a-Bloch) ADW profile for ($\xi_a=0,\pi$) and ($\xi_a=\frac{\pi}{2}$,$\frac{3\pi}{2}$), respectively. For a linear ADW profile centred at $x=0$ along the $x$-axis, $\dmxy = m_{\Delta}\big(\frac{\pi}{w}\big)\sin(\frac{\pi x}{w})\sin(\xi_a)$ for $|x| \leq \frac{w}{2}$, and zero elsewhere \cite{Jani2021} (see S5.1).  In other words, we expect ADWs to display a sinusoidal profile in $\dmxy$ ~and hence in $B_z$, with zero crossing at the centre and amplitude and sign modulated by $\sin{(\xi_a)}$. In particular, an a-N\'eel ADW will not yield a $B_z$ signal as $\dmxy = 0$, whereas an a-Bloch counterpart will show maximal signal. Based on Eq. \ref{eq:bz_realspace}, these characteristics are reflected in the simulated $B_z$ image of an ADW in Fig. \ref{fig:field}(a) (see S5.4) assuming a constant phase $\xi_a = \pi$. The measured $B_z$ image in Fig. \ref{fig:intro}(g), as well as a close-up image in Fig. \ref{fig:field}(b), capture the signature zero crossing of an ADW. DQM also reveals variations of $B_z$ along the wall boundary capturing the spatially varying nature of the phase $\xi_a$, as expected in \feo~\cite{Jani2021}. In effect, the full phase dependence of $B_z$ allows us to identify unambiguously the zero-signal sections along the varying $\xi_a$ wall boundary as a-N\'eel ADWs. Figure \ref{fig:field}(c) presents a reconstruction of the 
\mxy~ distribution of the multi-chiral ADW (illustrated by arrows), obtained by fitting its $B_z$ field to the experimental data in Fig. \ref{fig:field}(b) through systematic regularization strategies (see S7).

In contrast, DQM at $300~\mathrm{K}$ captures relatively larger spatial features of strong $B_z$ signal, as showcased in Fig. \ref{fig:intro}(h). This above-$T_{\mathrm{M}}$ regime is where we expect finite net magnetisation forming a tapestry of localised whirling topological structures, such as multi-chiral merons, antimerons and bimerons\cite{Jani2021}, as well as topologically-trivial in-plane domain walls. Each of these AFM textures produces a distinctive $B_z$ signal, allowing us to develop a systematic procedure to disambiguate them, see S5. Here we focus on two-dimensional (2D) topological textures and model isolated merons and antimerons based on a linear ansatz \cite{Jani2021,jacopo_2020}, described by the phase $\xi_a$ and the topological winding number $\mathcal{N}$ (see S5.3).  The magnetisation divergence for such an object in polar coordinates ($r,~\phi$) is $\dmxy = m_{\Delta}\sin\big(\phi (1-\mathcal{N})-\xi_a\big)f(r)$, where $f(r)$ is a radial function dependent on the (anti)meron phase (see S5.3). A meron ($\mathcal{N} = +1$) produces a radially symmetric $B_z$ distribution about its core with magnitude and polarity controlled by $\sin{(\xi_a)}$. Analogous to ADWs, a-N\'eel merons ($\xi_a=0$, $\pi$) are divergence-free and exhibit $B_z=0$, whereas a-Bloch counterparts ($\xi_a=\frac{\pi}{2}$, $\frac{3\pi}{2}$) show maximal $B_z$ amplitude. In contrast, for an antimeron ($\mathcal{N} = -1$), the $B_z$ distribution is two-fold symmetric and  $\xi_a$ controls a trivial azimuthal offset. Simulated $B_z$ images in Fig. \ref{fig:field}(d, g, j) of the a-Bloch meron of both polarities and the antimeron reinforce these observations (see S5.4). Thus, DQM unambiguously reveals the topological winding number $\mathcal{N}$ and the staggered vorticity $\vec{\mathcal{V}}$ for each spin texture. The final attribute of such a spin texture, the sign of its topological charge, is however rendered indistinguishable by DQM due to the vanishing canted moment at its core. Figure \ref{fig:field} (panels e, h, k) presents the measured $B_z$ images of an anti-clockwise a-Bloch meron ($\mathcal{N} = +1, \xi_a = \pi/2$), a clockwise  a-Bloch meron ($\mathcal{N} = +1, \xi_a = 3\pi/2$), and an antimeron ($\mathcal{N} = -1$), respectively, in good agreement with their simulated counterparts. Further, in  Fig. \ref{fig:field} (f, i, l) we once again reconstruct the corresponding \mxy~ distributions and use them to simulate the measured $B_z$ image for each texture discussed above (see S7). Given the typical density of spin textures evident in Fig. \ref{fig:intro}(h), the reconstructed \mxy~ approach better captures the finer details of the measured $B_z$ images in the absence of true isolation. We note that the initial density of (anti)merons can be reduced via meron-antimeron annihilation mediated by an external in-plane magnetic field  \cite{Jani2021} (see S8).
\begin{figure*}
    \centering
    \includegraphics[width =  1\textwidth]{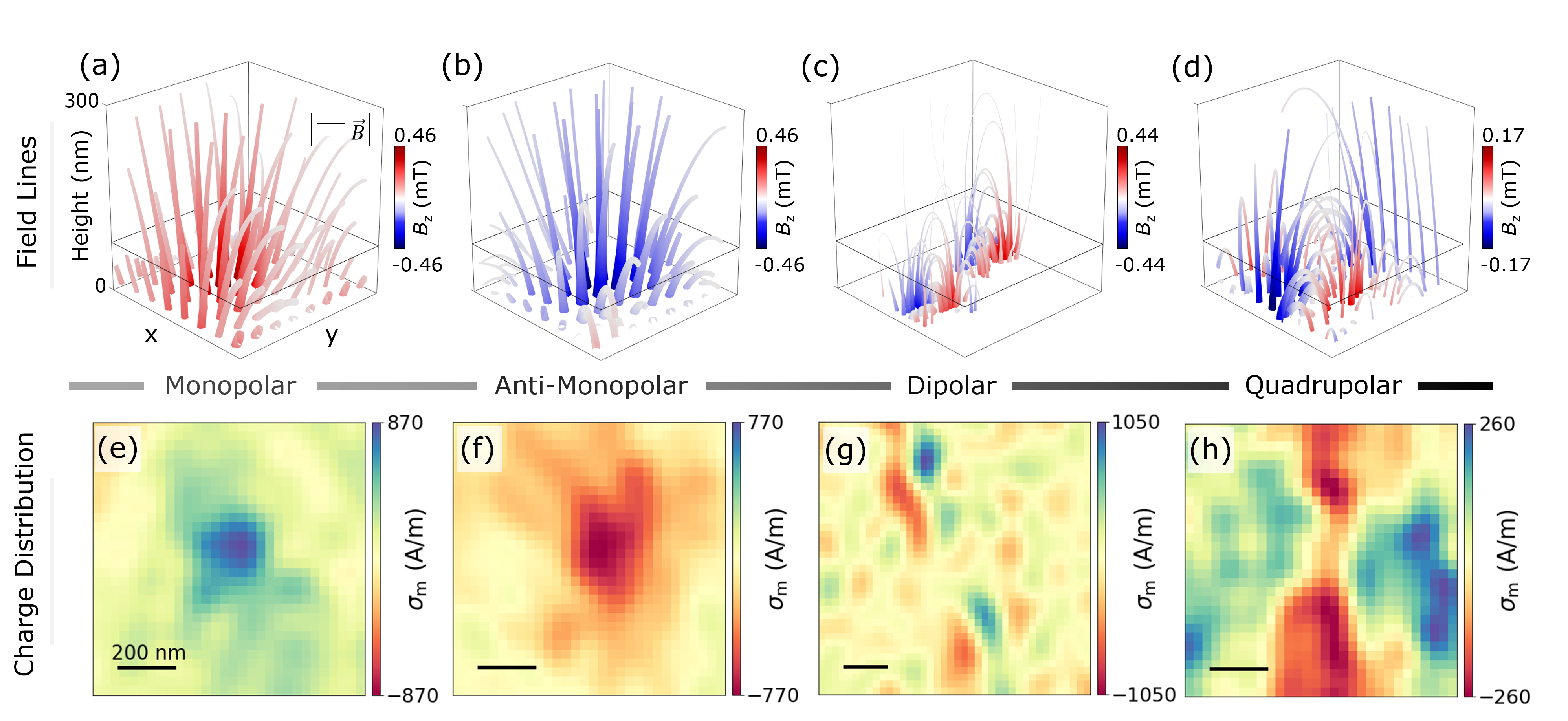}
    \caption{ \textbf{Emergent magnetic charge distributions. (a-d)}  Three-dimensional visualisation of $\vec{B}$ in the volume above an assortment of topological AFM textures. Streamtubes illustrate the magnetic field lines of $\vec{B}$ above (a) an anti-clockwise a-Bloch meron, (b) a clockwise a-Bloch meron, (c) an ADW and (d) an antimeron. The girth and colour of each streamtube vary with the magnetic field norm $|B|$ and the z-component of the field $B_z$, respectively. \textbf{(e-h)} Magnetic charge density $\sigma_m$ distributions retrieved from downward continuation of (a-d) reveal a magnetic monopolar, anti-monopolar, dipolar and quadrupolar charge character associated to an anti-clockwise a-Bloch meron, a clockwise a-Bloch meron, an ADW and an antimeron respectively. Scale bars are $200~\mathrm{nm}$.}
    \label{fig:charge}
\end{figure*}
Finally, a meron and an antimeron in close proximity can also form a stable bimeron. Figure \ref{fig:field}(m) displays the corresponding simulated $B_z$ image of an isolated bimeron, while Fig. \ref{fig:field}(n) shows the measured $B_z$ image of one such occurrence. As before, Fig. \ref{fig:field}(o) displays the simulated $B_z$ image from its reconstructed \mxy. It is interesting to note that while (anti)merons are always topologically protected, this is not necessarily true for meron-antimeron pairs. Labelling a meron-antimeron pair as topologically protected would require the knowledge of the sign of the topological charge of its constituents \cite{Galkina2010,Thoma2021}.

\section*{Emergent Magnetic Charge}
Reconstructing the \mxy~distribution from $B_z$ images is useful in understanding the underlying spin textures, but the fact that DQM actually provides a direct measure of \dmxy~creates a unique opportunity to consider a magnetic analog of the electric Gauss's law. Namely, the non-zero divergence of the magnetisation manifests the existence of an areal 2D magnetic charge density via $\sigma_m =- t (\dmxy)$. Here, $t$ is the film thickness and \mxy~is independent of $t$ -- valid in the thin film limit \cite{Jani2021}. Therefore, AFM textures in \feo~ have associated \emph{emergent magnetic charge} distributions, which locally act as sources or sinks of the magnetic field. In fact, we can define a formal \emph{duality relation} that connects the 2D magnetic charge density $\sigma_m$ to the staggered vorticity $\vec{\mathcal{V}}$ via
\begin{equation}
    \sigma_m/t  = - \dmxy = \Delta(\hat{z} \cdot \mathcal{\vec{V}}),
\label{eq:duality}
\end{equation}
which scales with $\sin(\xi_a)$, highlighting the influence of the texture phase. Crucially, retrieving the full 2D emergent charge density $\sigma_m$ only requires the Fourier deconvolution of the measured $B_z$ images from the transfer function $\alpha_{xy}$ (see S4.2). We can also perform a downward (upward) continuation \cite{blakely_1995} (see S4.2) of the planar $\vec{B}$ distribution captured in Fig. \ref{fig:field} (b, e, h, k, n), down to (away from) the sample surface. This allows a three-dimensional visualisation of $\vec{B}$ ($=\vec{H}$ \textit{in vacuo}) in the volume above the emergent magnetic charge distribution.

Figure \ref{fig:charge} panels (a) to (d) illustrate the representative field lines of $\vec{B}$ above a (anti-)clockwise a-Bloch meron, an ADW and an antimeron, respectively. For the two Bloch merons, $\vec{B}$ is consistent with the profile expected for spatially extended sources and sinks of magnetic field emanating from a \emph{monopolar} distribution, which we retrieved via Eq. \ref{eq:duality} in panels (e) and (f). This implies that a-Bloch merons host a class of emergent monopoles piggybacking on the whirling AFM textures. Interestingly, such monopolar magnetic distributions are not observed in ferromagnetic materials, as the presence of long-range demagnetizing fields favours divergence-free Bloch textures (e.g. in FM vortices). In our case, we are freed from this constraint due the presence of a weak demagnetizing contribution relative to the much stronger AFM exchange in \feo. In contrast to merons, the ADW in Fig. \ref{fig:charge}(c) and the antimeron in Fig. \ref{fig:charge}(d) are associated with $\sigma_m$ distributions that exhibit \emph{dipolar} and \emph{quadrupolar} characters, respectively, as illustrated in panels (g) and (h). Finally, we emphasise that the observation of emergent monopoles is fully consistent with the simulations of AFM topological textures in Fig. \ref{fig:field} and does not violate Maxwell's equation, $\vec{\nabla} \cdot \vec{B} = 0$.

\begin{figure*}
    \centering
    \includegraphics[width =  0.6\textwidth]{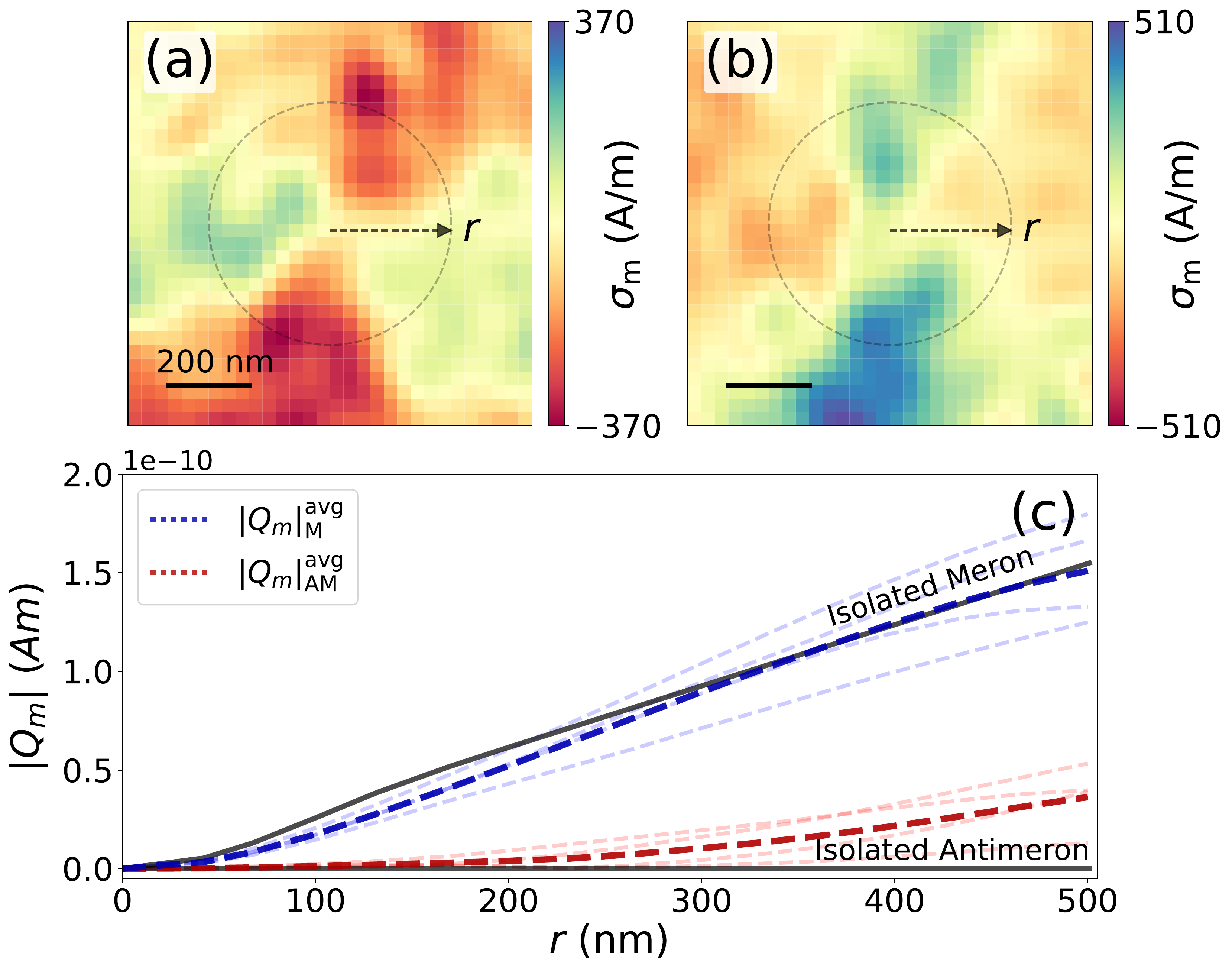}
    \caption{\textbf{Scaling of two-dimensional integrated magnetic charges. (a-b)} Reconstructed magnetic charge distribution, $\sigma_m$, of two experimentally observed antimerons (AM 1 and AM 2) with slightly distorted quadrupolar characters. AM 1 and AM 2 display a respective bias towards positive and negative charges, respectively. Dashed circle in (a) and (b) illustrates the circular integration area $S$ of radius $r$, centred at the core of the antimeron, to obtain $|Q_m|$. Scale bars are $200~\mathrm{nm}$. \textbf{(c)} Experimentally retrieved magnitude of the total integrated magnetic charge, $|Q_m|$ of multiple merons ($|Q_m|_\mathrm{M}$, light blue dashed curves) and antimerons ($|Q_m|_\mathrm{AM}$, light red dashed curves) plotted as a function of integration radius, $r$. Their average experimental $|Q_m|$ profiles, $|Q_m|^\mathrm{avg}_\mathrm{M}$ and $|Q_m|^\mathrm{avg}_\mathrm{AM}$, are represented by dark blue and dark red curves, respectively. Solid black curves plot the theoretically predicted $Q_m$ radial dependence based on Eq. \ref{meq:total_charge} for isolated merons and antimerons.}
    \label{fig:distort}
\end{figure*}

On the basis of the above analysis, it is tempting to attribute to isolated a-Bloch merons a non-zero net \emph{monopolar charge} quantified by $Q_m \equiv \int_S \sigma_m ~ds$ within an area $S$. We pick a circular integration area $S$ of radius $r$ centred on a given spin texture, as shown in Fig. \ref{fig:distort}(a and b) as an example in the case of antimerons. The particular $(1/r)$ dependence of $\dmxy$ for two-dimensional magnetic charges hosted by spin textures then yields (see S5.9)
\begin{align}
\scriptsize
\label{meq:total_charge}
    Q_m (r) = 
    \begin{cases} 
    ~ 2\pi~ m_\Delta~\sin{(\xi_a)}~\sin{\big(\frac{\pi r}{2 R_M} \big)}~r~t &,~\mathcal{N} = +1,~r\leq R_M\\
    ~ 2\pi~ m_\Delta~\sin{(\xi_a)}~r~t&,~\mathcal{N} = +1,~r> R_M\\
    ~0~&,~\mathcal{N} = -1
    \end{cases}~.
\end{align}
Figure \ref{fig:distort}(c) presents the radial dependence of $Q_m$ for four measured merons (light blue dashed curves) and antimerons (light red dashed curves), while the dark blue (red) dashed curve is the average $Q_m$ radial dependence for merons (antimerons). For an isolated linear meron model, $Q_m$ scales linearly with $r$, and the measured $Q_m$ radial dependence is in agreement with this predicted behaviour. $Q_m$ itself is not a topological invariant, as a smooth transformation of an a-Bloch meron to an a-N\'eel meron would tune $Q_m$ from non-zero to zero value. For an isolated antimeron, the two-fold rotational symmetry should ensure $Q_m = 0$ for all $r$ values (see S5.9) and at short distances the experimentally obtained $Q_m$ value of an antimeron agrees well with this prediction. Beyond typical measurement-induced variations, deviation from the strict $Q_m=0$ condition arises when an antimeron is influenced by neighbouring spin textures. This reduces the two-fold symmetry and reveals finite $Q_m$. Figures \ref{fig:distort}(a) and \ref{fig:distort}(b) capture this reduced symmetry and the consequential bias towards negative and positive $Q_m$, respectively. Finally, as $r$ goes to infinity, the integration area starts would overlap with surrounding spin textures, which leads to a further deviation from the $Q_m =0$ condition for an isolated antimeron model. 

The assumption that we have a collection of isolated spin textures somewhat oversimplifies the reality. While the AFM topological textures are mesoscopically discernible via DQM and X-ray techniques and thus appear localised, they are in fact the constituents of the complex multi-textural ensemble that interacts via a 2D magnetic charge canvas. The magnetic charge per constituent is therefore not just dictated by their nature as merons and antimerons, but modified through their interaction with other constituents in the ensemble. For example, an isolated meron and antimeron pair forming an AFM bimeron, as shown in Fig. \ref{fig:field}(n), would have a non-zero $Q_m$, whose sign is determined predominantly by its meron (see S5.9). However, this clearly cannot be the case for a bimeron embedded in a uniform magnetisation field in the far field (see S5.10), since the divergence theorem ensures that $Q_m=0$.  This goes to show that the interaction among AFM textures produces additional magnetic charge density away from the cores and highlights the interacting nature of this multi-textural ensemble.

\section*{Discussion and Outlook}
Our ability to identify the duality between topological AFM textures and magnetic charges  is due to the direct readout of staggered vorticity enabled by DQM, as the NV centre senses both amplitude and orientation of the weak magnetic field. As such, we go beyond detecting antimerons and merons, to further distinguish between clockwise and anti-clockwise a-Bloch components, which otherwise appear indistinguishable in X-ray dichroic images. Our imaging approach can be easily extended to a wider family of topological textures, including a-N\'eel merons and bimerons, skyrmions, and distorted AFM textures that are otherwise divergence-free (see S5.6-5.8), relevant for topological AFM circuitry \cite{AFM_review,Barker2016,oxide_skyrmions}. 

Although hematite provides especially favourable conditions for DQM imaging due to DMI-induced spin canting, it is by no means unique among AFMs in possessing a weak net magnetisation or quasi-isotropic spins in 2D \cite{Dzyaloshinsky1958,Moriya1960,FeBO3,orthoferrites}. In fully compensated antiferromagnets without bulk DMI, staggered spin textures can also generate local net magnetisation, either statically or dynamically \cite{tveten2016intrinsic,komineas2020traveling,AFM_skyrmion_motion}.  Moreover, the DQM technique can be useful in detecting preferential vorticity in ultra-thin films induced by interfacial interactions –– a key requirement for applications of quantum materials in topological spintronics \cite{Fe2O3_dmi,oxide_skyrmions}. 
 
In light of our discovery of the duality between magnetic charges and topological AFM textures, it is tempting to compare hematite with other systems that harbour emergent magnetic monopoles, such as the pyrochlore spin ice \cite{Castelnovo2008}.
Although intriguing, monopoles in spin ice are intrinsically distinct, as they have an underlying gauge charge, which is topological and quantised. 
On the contrary, the emergent magnetic charges in hametite are two-dimensional, not quantised and are only topological in the sense that they dress topological AFM textures underpinning them. In fact, we have demonstrated that hematite supports a rich tapestry of interacting magnetic charge distributions which could open up new and complementary ways to detect, manipulate and functionalise -- via their magnetic charge -- AFM topological textures. For example, our capability to classify different AFM spin textures could be combined with conventional spin manipulation techniques such as various spin torques, allowing for natural read-out and write-in schemes. Realizing this in a complex manifold of states endowed with highly non-linear interactions could be attractive for unconventional computing \cite{Prychynenko2017,bourianoff2018potential}. Finally, the intriguing physical insights revealed in \feo~are a testament to the relevance and potential of DQM as a versatile table-top platform to explore emergent phenomena in antiferromagnets and other quantum materials.

\section{Acknowledgements}
We like to thank D. M. Kara, A. Soumyanarayanan and J. J. Robinson for their useful discussions and valuable insights. Work performed at the University of Cambridge was supported by the Cambridge Nanoscale Sensing and Imaging Suite (CANSIS) as part of the Henry Royce Institute under Engineering and Physical Sciences Research Council (EPSRC) grant No. EP/P024947/1. We further acknowledge funding from EPSRC QUES2T (EP/N015118/1) and the Betty and Gordon Moore Foundation. Work done at the National University of Singapore was supported by the Agency for Science Technology \& Research (A*STAR) under Advanced Manufacturing \& Engineering Individual Research Grant (A1983c0034). Work done at the University of Oxford was supported by EPSRC grant (EP/M020517/1) and the Oxford-ShanghaiTech collaboration project. A.K.C.T. acknowledges funding from A*STAR, through the National Science Scholarship. H.J. acknowledges the support of Marie Skłodowska-Curie Postdoctoral Fellowship under the Horizon Europe Guarantee Funding from UK Research and Innovation (EP/X024938/1). M.H. acknowledges funding from EPSRC NQIT (EP/M013243/1). L.S. acknowledges funding from the European Union’s Horizon 2020 research and innovation programme under the Marie Skłodowska-Curie grant agreement No. 101063285. H.S.K. acknowledges funding from the Royal Society through a University Research Fellowship.

\newpage\hbox{}\thispagestyle{empty}\newpage
\onecolumngrid
\appendix

\makeatletter 
\renewcommand{\thefigure}{S\@arabic\c@figure}
\makeatother

\makeatletter 
\renewcommand{\thetable}{S\@arabic\c@table}
\makeatother

\makeatletter 
\renewcommand{\thesection}{\Alph{section}}
\makeatother

\makeatletter 
\renewcommand{\thesubsection}{\thesection\arabic{subsection}}
\makeatother

\section{Material Growth and Characterization}
\label{si-sec:material_character}
\begin{figure}[H]
    \centering
    \includegraphics[width =  0.4\textwidth]{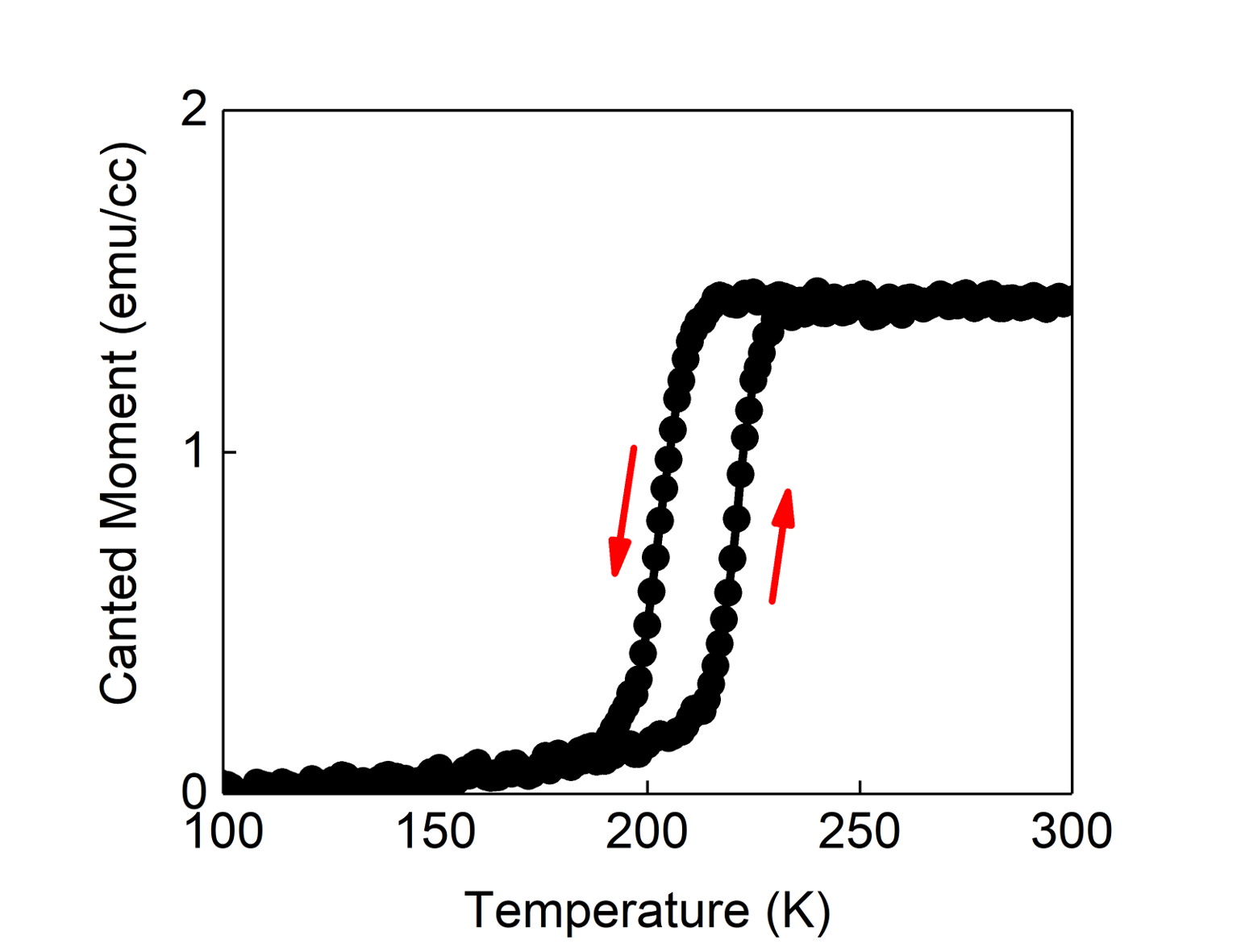}
    \caption{ \textbf{Magnetometry characterisation.} Temperature dependence characterisation (M-T) with an in-situ 50 mT in-plane field. Arrows indicate the direction of the temperature sweep.}
    \label{sfig:magnetometry}
\end{figure}

Epitaxial \feo~films were pulsed-laser deposited on (0001)-oriented single crystalline \ensuremath{\alpha}-Al\ensuremath{_2}O\ensuremath{_3} substrates (CrysTec GmbH), using a KrF excimer laser at 248 nm. X-ray reflectometry indicated a film thickness $t = 25~\mathrm{nm}$. A detailed fabrication recipe can be found in our previous studies\cite{Jani2021,Jani2021b}. No capping overlayer was used in this study. Bulk magnetometery was performed using a Superconducting Quantum Interference Device magnetometer (Quantum Design). The temperature dependence of the canted moment, shown in Fig. \ref{sfig:magnetometry}, reveals a Morin transition in the range of $200-220~\mathrm{K}$ such that the magneto-crystalline anisotropy changes from in-plane ($T>T_M$) to out-of-plane ($T<T_M$). 
\section{Setup and Imaging Details}
The diamond quantum magnetometry (DQM) setup consists of an electrical readout atomic force microscope in a closed-cycle cryostat (attoDRY1000, Attocube systems) integrated with a home-built confocal setup. For magnetic field sensing, CW-ODMR was employed for sensor characterisation, and switched to pulsed-ODMR for the imaging of \feo~to improve sensitivity. The diamond sensor used in our measurements (Qzabre AG) contains Nitrogen-15 ($I = 1/2$) thereby giving two hyperfine resonances per electron spin resonance with a 3 MHz splitting that is easily detectable with pulsed-ODMR. The on-axis field magnitude is determined from the ODMR frequency splitting via the weak field approximation\cite{Rondin2014}, where the influence of the off-axis field on the zero-field splitting is neglected. Microwave excitation is delivered via a copper wire pulled across the sample surface. Imaging is conducted with the diamond sensor positioned $\sim 50~\mathrm{\mu m}$ away from the wire antenna. A Tektronix AWG (70002A) was programmed to provide the pulse sequences necessary for pulsed-ODMR. Typical scans have a pixel size of $\sim 33.3~\nm$, with an averaging time per pixel of $50-100~\mathrm{\mu s}$ at both 4 K and 300 K. The total time for a $5 \times 5~\mathrm{\mu m^2}$ image is approximately twelve hours, with the retrace speed being $\leq 0.1~\mathrm{Hz}$ to avoid physical damage to the sample surface by the diamond pillar.

\section{Sensor Characterisation}
\label{si-sec:nc_character}
\subsection{NV-axis Orientation}
\begin{figure}
    \centering
    \includegraphics[width =  0.9\textwidth]{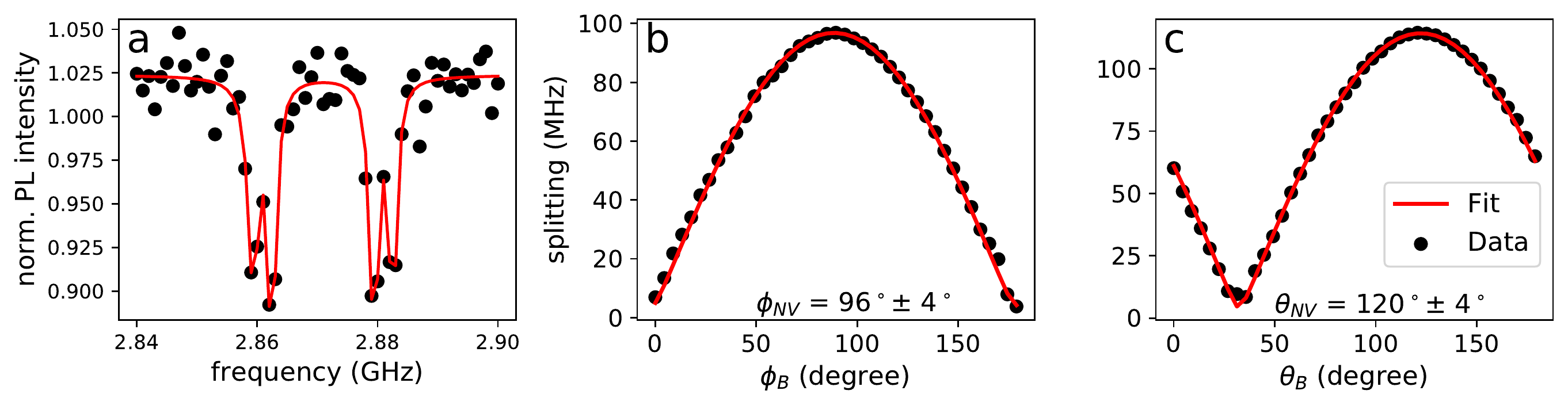}
    \caption{ \textbf{NV orientation characterisation.} (a) ODMR with 0.5 mT applied along the NV axis showing the Zeeman-split resonances with additional hyperfine splitting due to the host ${}^{15}$N. ODMR spectrum as a function of (b)~$\phi_B$ with $\theta_B$ fixed at $180^\circ$ and (c)~$\theta_B$ with $\phi_B =96^\circ $. The maximum splitting in (b) determines $\phi_{NV} = 96 ^\circ$ which is then used in (c) to identify $\theta_{NV} = 120^\circ$.}
    \label{fig:nvaxis}
\end{figure}

NV-axis orientation with reference to the lab frame is characterized by applying an external magnetic field $\vec{B}(B_0,\theta_B,\phi_B)$ using a three-axis Helmholtz coil. We characterize the NV orientation by obtaining ODMR spectra while holding the $B$-field at a constant magnitude $B_0$ and polar $\theta_B$ (azimuthal $\phi_B$) angle and sweeping $\phi_B$ ($\theta_B$) (details are found in \cite{Stefan2021}). The NV-axis orientation characterisation of the diamond sensor used for imaging in Figs 1-3 in the main text (fig.\ref{fig:nvaxis}), shows $\theta_{NV} = 120\pm 4 ^{\circ}$ and a $\phi_{NV} = 96 \pm 4^{\circ}$.

\subsection{NV-to-Sample Distance}
\begin{figure}
    \centering
    \includegraphics[width =  0.9\textwidth]{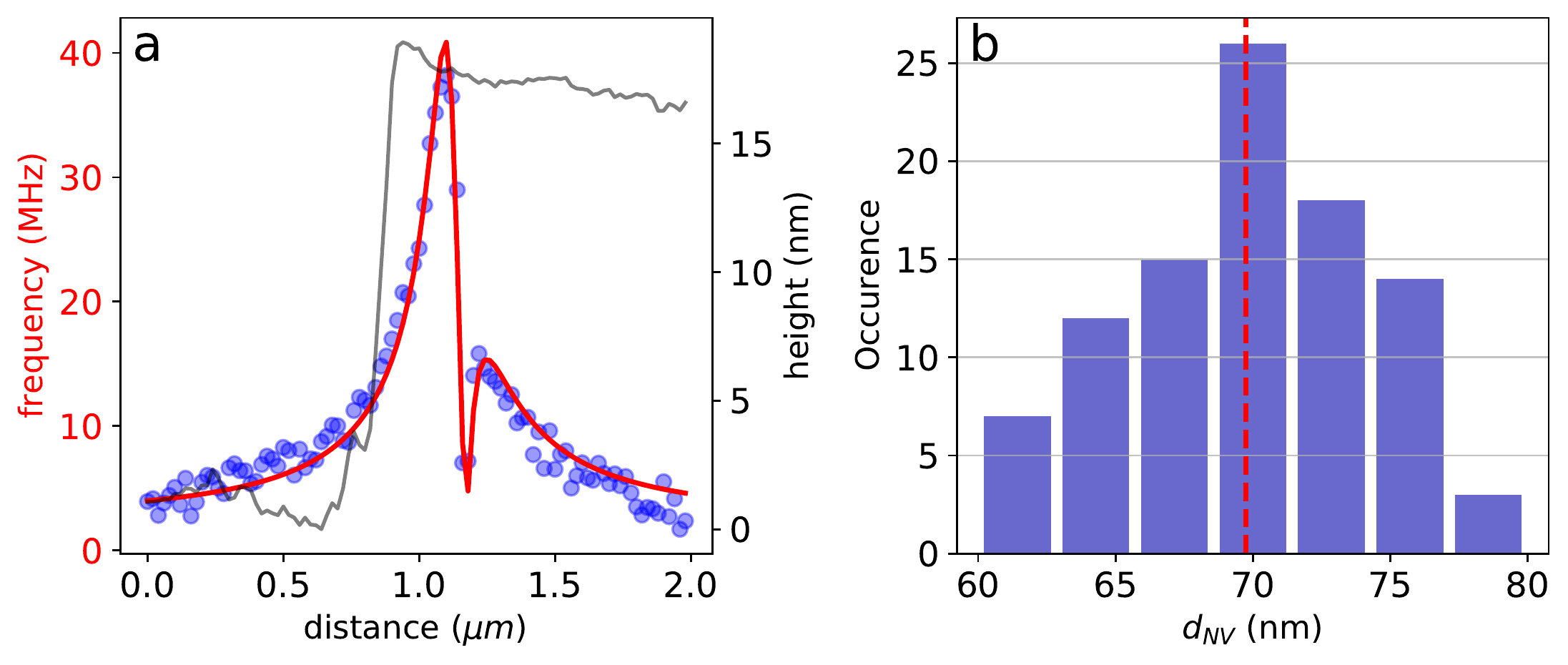}
    \caption{ \textbf{NV-to-sample distance characterisation.} (a) ODMR spectrum across the edge of a CoFeB magnetic strip plotted on top of the corresponding topography. From the fit one can extract the NV-to-sample distance. (b) Height distribution extracted from 100 line scans across the edge, giving an average NV-to-sample distance of $70\pm6~\nm$. }
    \label{fig:nvdist}
\end{figure}

The NV-to-sample distance $d_{NV}$ is determined by measuring the magnetic field emitted across the edge of an out-of-plane magnetic material with the sensor of interest. Our calibration sample is a CoFeB magnetic strip (details in \cite{Stefan2021}) with a magnetization that remains saturated at remanence. This allows for accurate $d_{NV}$ value extraction following the proposal in \cite{Hingant2015}. An example ODMR linescan across the edge of the magnetic strip, obtained with the diamond sensor used for imaging in Figs. 1-3 in the main text, is given in Fig. \ref{fig:nvdist} (a). A distribution of extracted $d_{NV}$ values from multiple ODMR linescans, such as the one in (a), is given in Figure \ref{fig:nvdist} (b), resulting in an average $d_{NV}$ value of $70 \pm6~ \nm$ which sets the effective spatial resolution of our imaging. This is sufficient to study the AFM topological textures and piggybacked charges in the study, whose length scales are larger.

\section{Magnetic Field Analysis}
\label{si-sec:field_analysis}
\subsection{Magnetic Field Components Above Sample Surface}
\label{si-sec:diptensor}
Assuming a density of magnetic dipoles $\vec{m}(x,y)$ of thickness \textit{t}, the magnetic field $\vec{B}(x,y)$ generated at a distance \textit{d} above the sample surface (source-free region) can be described in Fourier space via the dipolar tensor \cite{Harte1968,VanDerSar2015,Lima2009}:
\begin{equation}
\begin{pmatrix}
    \tilde{B}_x(k_x,k_y,d)\\
    \tilde{B}_y(k_x,k_y,d)\\
    \tilde{B}_z(k_x,k_y,d) 
\end{pmatrix} = - \frac{\mu_0}{2}e^{-kd}(1-e^{-kt})
\begin{pmatrix}
    k_x^2/k^2 & k_xk_y/k^2 & ik_x/k \\
    k_xk_y/k^2 & k_y^2/k^2 & ik_y/k\\
    ik_x/k &  ik_y/k & -1
\end{pmatrix}
\begin{pmatrix}
    \tilde{m}_x(k_x,k_y)\\
    \tilde{m}_y(k_x,k_y)\\
    \tilde{m}_z(k_x,k_y) 
\end{pmatrix}.
\label{seq:diptensor}\end{equation}
In the absence of a canted $m_z$ component in $\alpha$-Fe$_{2}$O$_{3}$, we have:
\begin{equation}
     \tilde{B}_z(k_x,k_y,d) = - \frac{\mu_0}{2}e^{-kd}(1-e^{-kt})\left(\frac{ik_x \tilde{m}_x+ik_y \tilde{m}_y}{k}\right)
     \label{seq:bz_fourier},
\end{equation}
\begin{equation}
    B_z = \mathscr{F}^{-1}\left[\left(-\frac{\mu_0 e^{-kd}(1-e^{-kt})}{2k}\right)(ik_x \tilde{m}_x+ik_y \tilde{m}_y)\right]= \alpha_{xy}(d,t) * \vec{\nabla} \cdot \mxy .
    \label{seq:bz_real}
\end{equation}
In equation \ref{seq:bz_real} we use the convolution theorem, and define $\alpha_{xy}(d,t) \equiv \mathscr{F}^{-1}(-\frac{\mu_0 e^{-kd}(1-e^{-kt})}{2k})$ and $\mathscr{F}$ and $\mathscr{F}^{-1}$ are the Fourier transform and its inverse. In real space, one can intuitively interpret equation \ref{seq:bz_real} as the convolution of a point spread function, $\alpha_{xy}$\cite{Dovzhenko2018}, with the divergence of the in-plane canted magnetization, \dmxy. In fact, as the distance $d$ above the sample surface increases, spatial features of \dmxy~get increasingly `blurred out', decreasing the effective spatial resolution of the NV sensor. A general discussion involving all components of the magnetization can be found in \cite{Dovzhenko2018}.
\subsection{Downward (Upward) Continuation and Charge Retrieval}
\label{si-sec:continuation}
Given the constrains set out in section \ref{si-sec:diptensor} and following equation \ref{seq:diptensor}, one can write the relation \cite{Thiel2019,blakely_1995}:
\begin{equation}
    \vec{\tilde{B}}(d+\Delta d) = e^{-k \Delta d} \vec{\tilde{B}}(d) ,
    \label{seq:continuation} 
\end{equation}
where $\vec{\tilde{B}} = \mathscr{F}(\vec{B})$. A positive or negative $\Delta d$ in equation \ref{seq:continuation} corresponds to an \textit{upward} or \textit{downward} continuation of the $B$-field distribution captured at $d$. Notice that when $\Delta d$ is positive, the exponential term $e^{-k \Delta d}$ acts to smooth out high frequency components and physically translates to losing spatial resolution. In the negative case, the exponential term tends to  amplify high frequency noise present in the  measurements. For this reason, we apply a Hanning window with a cut-off frequency at $2\pi/d_{NV}$, which is a common method to suppress the undesired amplification of high-frequency noise \cite{Thiel2019,Wornle2021,marchiori2022nanoscale}.

By applying the \textit{downward continuation} method down to the sample surface, it is therefore possible to retrieve the $\sigma_m= -t~\dmxy$ given by:
\begin{equation}
    \sigma_m= -t~\mathscr{F}^{-1}\big( \tilde{B}_z / \tilde{\alpha}_{xy}(t,d=d_\mathrm{NV})\big) ,
\end{equation}
where $\tilde{\alpha}_{xy} = \mathscr{F}(\alpha_{xy})$. This is a deconvolution process with the function $\alpha_{xy}$.
 
\section{Magnetic Models}
\label{si-sec:linear_model}
The presence of $\vec{\mathcal{D}}$ along the $c$-axis in \feo~(bulk DMI) produces a small in-plane canting of angle $\Delta$ in the basal $ab$-plane, generating a net magnetization which can be obtained from the local N\'eel vector using $\vec{m}=\vec{\mathcal{D}}\times \vec{l} = \Delta \hat{\eta}\times \vec{l}$~ in the small angle approximation. In \feo, the unit vector $\hat{\eta} = (0, 0, +1)$\cite{Thoma2021} is parallel to the \textit{c}-axis. Since in our case the \textit{c}-axis is along the $z$-axis of the lab frame, the expression is simplified to $\vec{m}= \Delta \hat{z}\times \vec{l}$. Note that this condition is strictly enforced in the trigonal symmetry of the paramagnetic phase but is relaxed at lower temperatures due to magneto-elastic interactions, resulting in a small out-of-plane moment, which has hence been ignored here  \cite{Dzyaloshinsky1958}. With this, we will now describe the spatial distributions of the canted moment in various AFM quasiparticles considering their respective linear ansatz.

Herein, we assume a canted moment $m_\Delta \sim 2 \times 10^3~\mathrm{A/m}$, and a canting angle $\Delta \sim 1.1~\mathrm{mrad}$. This is consistent with values reported for bulk \feo~ (Table 4.2 of \cite{Morrish1995}) and comparable to the magnetic response of our samples, as discussed in our previous studies \cite{Jani2021,Jani2021b}. In the small angle approximation, the average magnitude of the sublattice magnetization $M_s = m_\Delta/(2 \Delta) \sim 9.1\times 10^6~\mathrm{A/m}$.

\subsection{Magnetic Field and Charge of AFM Antiphase Domain Walls}
Assuming a linear ansatz for the antiphase domain wall profile centered at $x=0$ along the $x$-axis, with phase $\xi_a$ and width $w$, we have:
\begin{align}
    \vec{l} &= 2 M_s~ \big(~\cos{\theta}\cos{\xi_a}~,~\cos{\theta}\sin{\xi_a}~,~\sin{\theta}~\big)~,\\
    \vec{m} &= 2 \Delta M_s~ \big(~-\cos{\theta}\sin{\xi_a}~,~\cos{\theta}\cos{\xi_a}~,~0~\big)
    \label{seq:m_adw} ~,
\end{align}
\begin{align*}
    \mathrm{where,~}
    ~~\theta = \begin{cases} 
    -\frac{\pi}{2} &,~x<-\frac{w}{2}\\
    \frac{\pi x}{w} &,~|x|\leq\frac{w}{2}\\
    +\frac{\pi}{2} &,~x>\frac{w}{2}\\
    \end{cases}~.
\end{align*}
Hence,
\begin{equation}
    \vec{\nabla} \cdot \vec{m} = 
    \begin{cases}
    2 \Delta M_s \big(\frac{\pi}{w}\big)  \sin{\theta} \sin{\xi_a}~&,~|x| \leq \frac{w}{2}\\
    0~&,~|x| > \frac{w}{2}
    \end{cases}~. 
\end{equation}
\subsection{Magnetic Field and Charge of AFM In-plane Domain Walls}
Uniform IP AFM domains do not generate any stray fields since their magnetisation would be non-divergent (see main text Eq. 1). However, in-plane domain walls above the Morin transition, which are expected in \feo to span $60^\circ$ and $120^\circ$, can generate stray fields. Their stray field signature depends on the domain wall width $w$, winding direction $l_0$, internal phase of the domain wall -- given by the angular span $\eta_0$, and phase offset $\phi_0$:
\begin{align}
    \vec{l} &= 2 M_s~ \big(~\cos{\big(\Psi+\phi_0\big)}~,~\sin{\big(\Psi+\phi_0}\big)~,~0~\big)~,\\
    \vec{m} &= 2 \Delta M_s~ \big(~-\sin{\big(\Psi+\phi_0\big)}~,~\cos{\big(\Psi+\phi_0\big)}~,~0~\big)
    \label{seq:m_ipdw} ~,
\end{align}
\begin{align*}
    \mathrm{where,~}
    ~~\Psi = \begin{cases} 
    -\frac{l_0\eta_0}{2} &,~x<-\frac{w}{2}\\
    \frac{l_0\eta_0 x}{w} &,~|x|\leq\frac{w}{2}\\
    +\frac{l_0\eta_0}{2} &,~x>\frac{w}{2}\\
    \end{cases}~.
\end{align*}
Hence,
\begin{equation}
    \vec{\nabla} \cdot \vec{m} = 
    \begin{cases}
    -2 \Delta M_s \big(\frac{l_0\eta_0}{w}\big)  \cos{\Psi + \phi_0} ~&,~|x| \leq \frac{w}{2}\\
    0~&,~|x| > \frac{w}{2}
    \end{cases}~. 
\end{equation}

IP domain walls generate different stray fields depending on their internal phase as illustrated in Fig. \ref{fig:sim_bz_ipdw}.

\subsection{Magnetic Field and Charge of AFM Merons and Antimerons}
\label{si-sec:tt}
Assuming a linear ansatz for an (anti)meron centered at the origin, with phase $\xi_a$, topological winding number $\mathcal{N}$, and radial size $R_M$, we have:
\begin{align}\label{seq:M_meron}
    \vec{l} &= 2 M_s~ \big(~\sin{\theta}\cos{(\mathcal{N}\phi + \xi_a)}~,~\sin{\theta}\sin{(\mathcal{N}\phi + \xi_a)}~,~ \cos{\theta}~\big)~,\\
    \vec{m} &= 2 \Delta M_s~ \big(~-\sin{\theta}\sin{(\mathcal{N}\phi + \xi_a)}~,~\sin{\theta}\cos{(\mathcal{N}\phi + \xi_a)}~,~0~\big)~,
    \label{seq:m_meron}
\end{align}
\begin{align*}
    \mathrm{where,~} 
    ~~\theta = \begin{cases} 
    \frac{\pi r}{2R_M} &,~r\leq R_M\\
    \frac{\pi}{2} &,~r>R_M\\
    \end{cases}~.
\end{align*}
Hence,
\begin{align}
    \vec{\nabla} \cdot\vec{m} = 
    \begin{cases}
        2\Delta M_s \sin{( (1-\mathcal{N})\phi - \xi_a)} \Big( \frac{\mathcal{N}}{r}\sin{\theta} + \frac{\pi}{2 R_M}\cos{\theta} \Big)&, r\leq R_M\\
        2 \Delta M_s \sin{( (1-\mathcal{N})\phi - \xi_a)} \Big( \frac{\mathcal{N}}{r} \Big)&, r > R_M
    \end{cases}~.
\end{align}
The last term is referred to as a radial function $f(r)$ in the main text.

\subsection{Magnetic Field Simulations of Various AFM Entities}
\label{si-sec:sims}
\begin{figure}
    \centering
    \includegraphics[width =  0.9\textwidth]{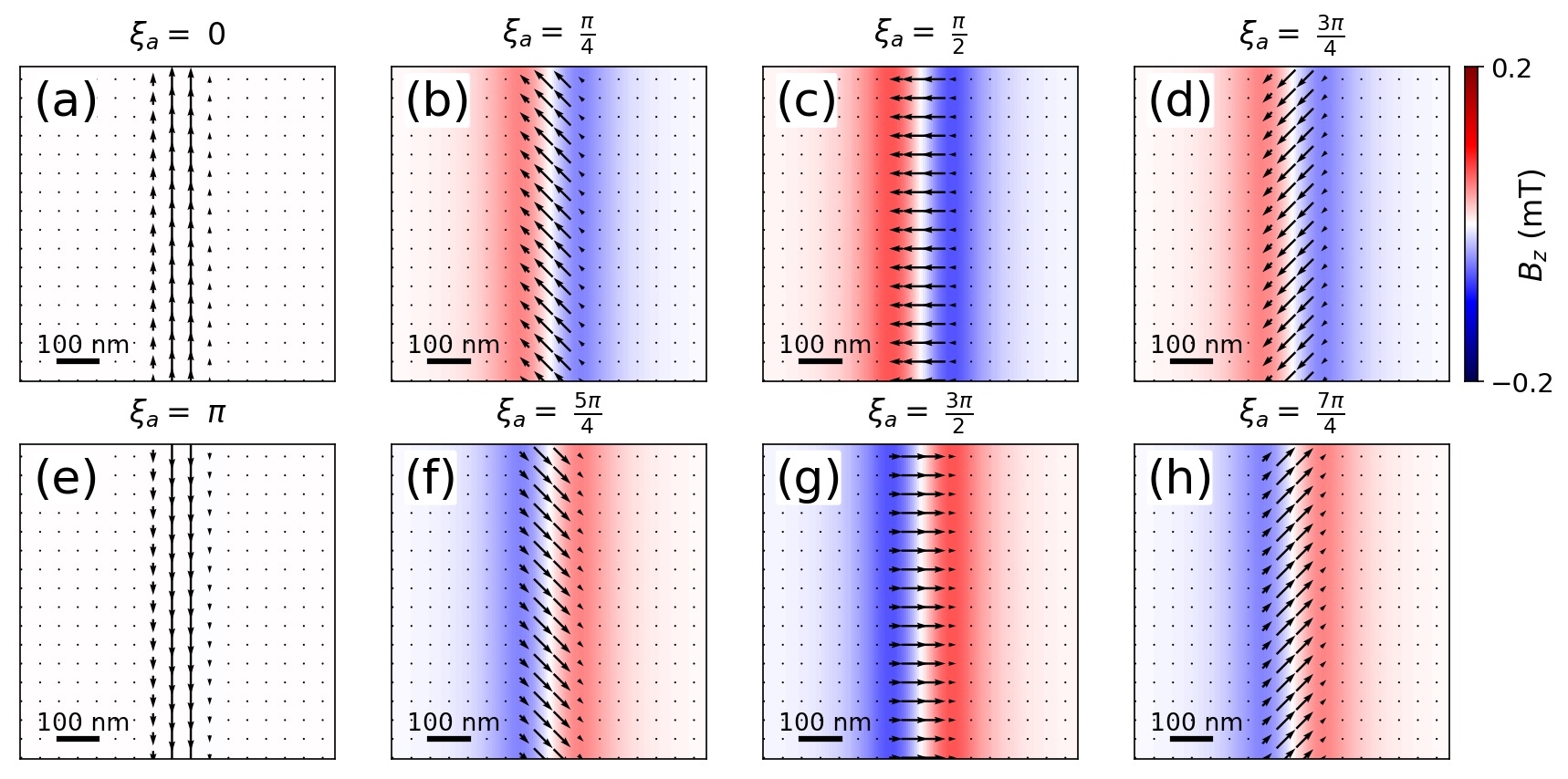}
    \caption{ \textbf{Magnetic field simulation of one-dimensional topological textures.} Simulated $B_z$ exhibited by ADWs upon varying $\xi_a$. Black arrows illustrate \mxy.}
    \label{fig:sim_bz_adw}
\end{figure}

\begin{figure}
    \centering
    \includegraphics[width =  0.9\textwidth]{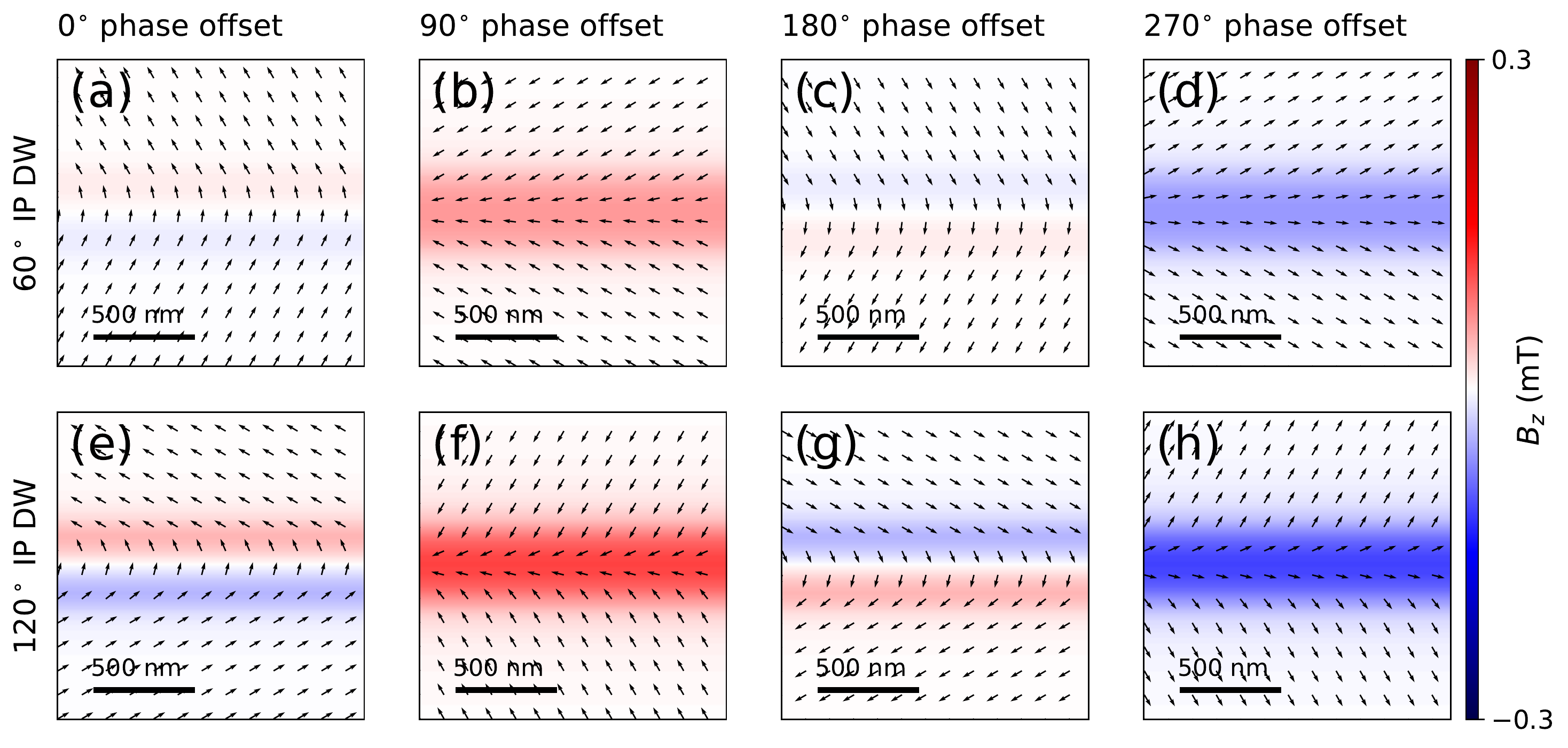}
    \caption{ \textbf{Magnetic field simulation of trivial in-plane domain walls.} Simulated $B_z$ exhibited by 60$^\circ$ and 120$^\circ$ in-plane domain walls with various phase offset. Black arrows illustrate \mxy.}
    \label{fig:sim_bz_ipdw}
\end{figure}

\begin{figure}
    \centering
    \includegraphics[width =  0.9\textwidth]{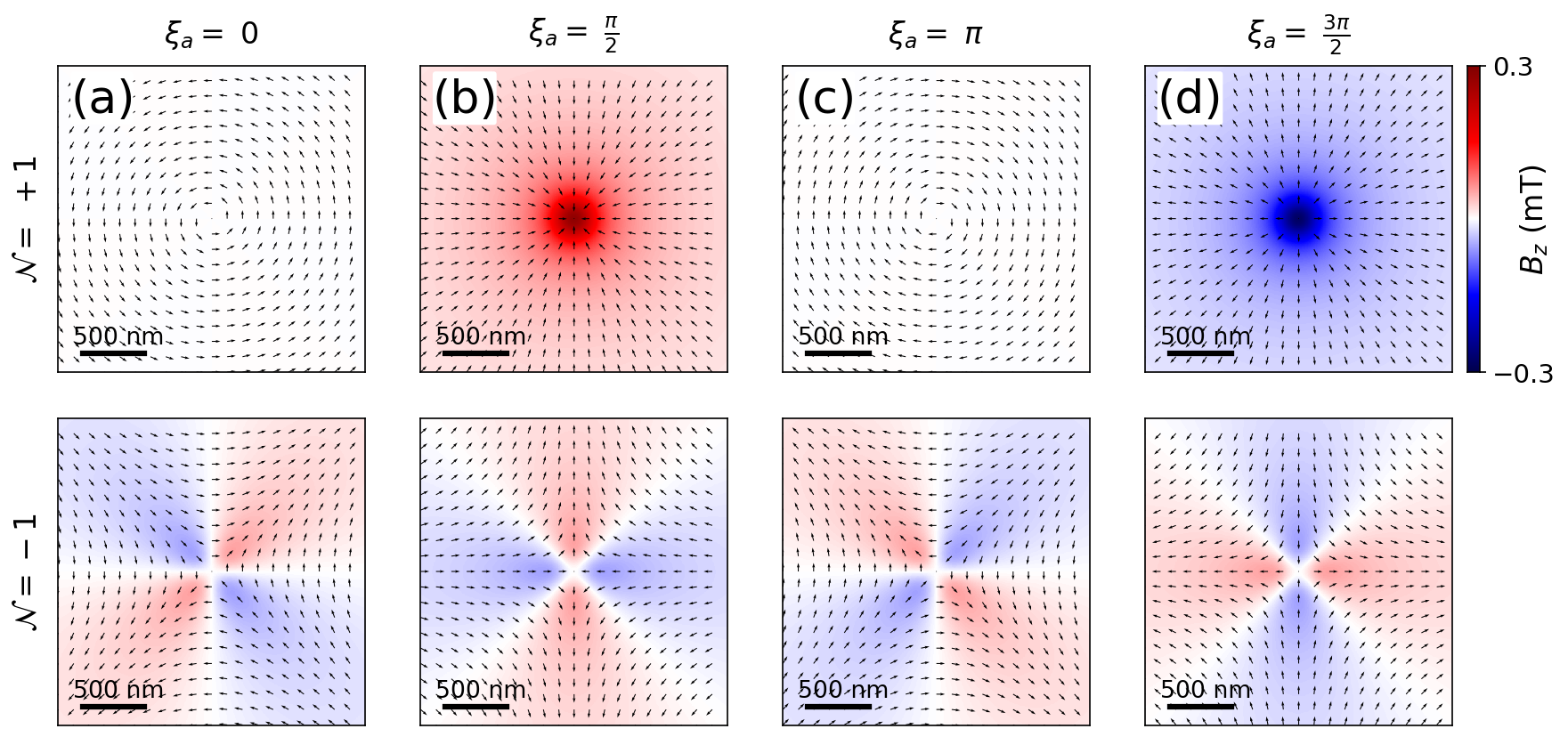}
    \caption{ \textbf{Magnetic field simulation of two-dimensional topological textures.} Simulated $B_z$ exhibited by merons ($\mathcal{N}=+1$) and antimerons ($\mathcal{N}=-1$) upon varying $\xi_a$. Black arrows illustrate \mxy.}
    \label{fig:sim_bz_quasi}
\end{figure}

We have simulated the magnetic fields generated $d=70~\nm$ above $\vec{m}$ grids containing antiphase domain walls in Fig. \ref{fig:sim_bz_adw}, $60^\circ$ and $120^\circ$ in-plane domain walls in Fig. \ref{fig:sim_bz_ipdw}, and merons and antimerons in Fig. \ref{fig:sim_bz_quasi}.
\subsection{Distinguishing Topological and non-topological AFM textures}
\label{si-sec:distinguish}
Above the Morin transition, topological textures are always intertwined with IP domain walls – in fact, each topological texture is a `pinch-point' of IP  domain walls \cite{Jani2021,Chmiel2018}. It is therefore crucial to be able distinguish them apart. 
Whilst IP domain walls can generate stray fields (fig. \ref{fig:sim_bz_ipdw}), their strength, distribution, and evolution are markedly different from counterparts generated by merons and antimerons (Fig. \ref{fig:sim_bz_quasi}). In the following we outline the salient differences and how one can distinguish them:
\begin{itemize}
\item Locally, topological merons and antimerons are two-dimensional entities which produce confined stray fields, characteristic of monopolar and quadrupolar distributions. Alternatively, IP domain walls are deconfined quasi-one-dimensional trivial texturesThis is illustrated in Fig. \ref{fig:linear_charge} using the linear meron ansatz. 
\begin{figure}
    \centering
    \includegraphics[width =  0.5\textwidth]{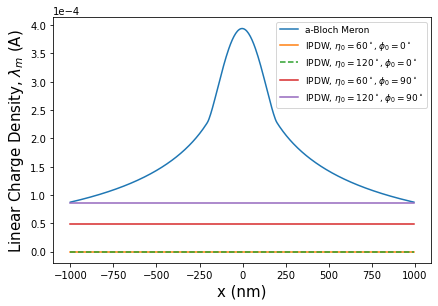}
    \caption{\textbf{Linear charge profile merons and IP domain walls.} The linear charge profile $\lambda_M$ is the integrated charge along the y-axis given as a function of the x-axis of the meron in Fig. \ref{fig:sim_bz_quasi} and the IP domain walls (IPDW) in Fig. \ref{fig:sim_bz_ipdw}. The $\lambda_M$ of the meron and IPDWs reveal their respective confined and deconfined nature.}
    \label{fig:linear_charge}
\end{figure}

\item The maximum stray field magnitude generated by merons and domain walls is (i) inversely proportional to their characteristic length scale (e.g., meron core radius; IP domain wall width, $W$) and (ii) directly proportional to the total angle they span in spin space ( $\pi$ for meron; $\eta_0$ for IP domain wall angle) and is given by:
\begin{gather}
    B^\mathrm{max}_{z, \mathrm{meron}} \propto |\dmxy|^\mathrm{max}_\mathrm{meron} = 2 \Delta M_s \left(\frac{\pi}{R_M}\right)\\
    B^\mathrm{max}_{z, \mathrm{IP wall}} \propto |\dmxy|^\mathrm{max}_\mathrm{IP wall} = 2 \Delta M_s \left(\frac{\eta_0}{W}\right)
\end{gather}
Since merons trap out-of-plane cores, the anisotropy energy penalty causes these cores to be small\cite{Jani2021}. On the other hand, IP domain walls are much wider as they do not have out-of-plane regions and are only limited by the very weak basal anisotropy in \feo \cite{Besser}. This causes $R_M < W$, at temperatures well above the Morin transition, which is the case at 300 K. Moreover, since IP domain walls are expected to have $\eta_0 \leq \pi$. Consequently, the magnitude of the stray field generated by a-Bloch merons at their core is significantly more intense, by a factor of $\sim2$ compared to trivial textures. This allows unambiguous identification of topological meron cores.
\item Although both topological and trivial textures are present above the Morin transition, application of in-plane magnetic fields results in the large-scale annihilation of merons and antimerons only. This triggers the selective elimination of strong and confined monopolar stray fields, leaving behind trivial textures, as illustrated in SI section \ref{si-sec:field_imaging} and our previous work\cite{Jani2021,Chmiel2018}. This experimentally evinces the different evolution of topological and non-topological entities.
\item As for antimerons, the quadrupolar stray field distribution with a zero-field cross pattern (Fig. \ref{fig:sim_bz_quasi}) while distinct from IP domain walls (Fig. \ref{fig:sim_bz_ipdw}), can be challenging to disambiguate in isolation. For this reason, we restrict antimeron identification and fitting to regions in the immediate proximity of strong monopolar field distributions generated by merons. This is justified, given that merons and antimerons occur with equal probability after executing a Kibble-Zurek transition\cite{Jani2021,Chmiel2018} and that, based on our X-ray work, meron-antimeron pairs are extremely common.          
\end{itemize}
\subsection{Invisibility of Divergence-free Textures}
\label{ssec:div-zero}
\begin{figure}
    \centering
    \includegraphics[width =  0.9\textwidth]{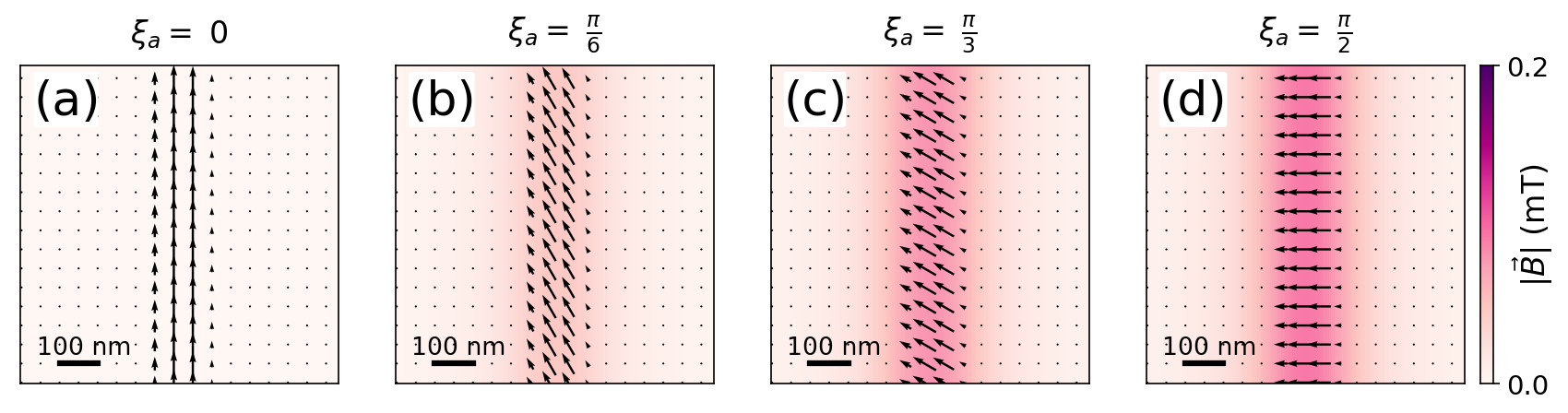}
    \caption{ \textbf{Field magnitude of ADWs with different phase $\xi_a$.} Field magnitude $|\vec{B}|$ obtained from the magnetic field simulations of ADWs with different $\xi_a$. Black arrows illustrate \mxy.}
    \label{fig:sim_b_mag_adw}
\end{figure}

\begin{figure}
    \centering
    \includegraphics[width =  0.9\textwidth]{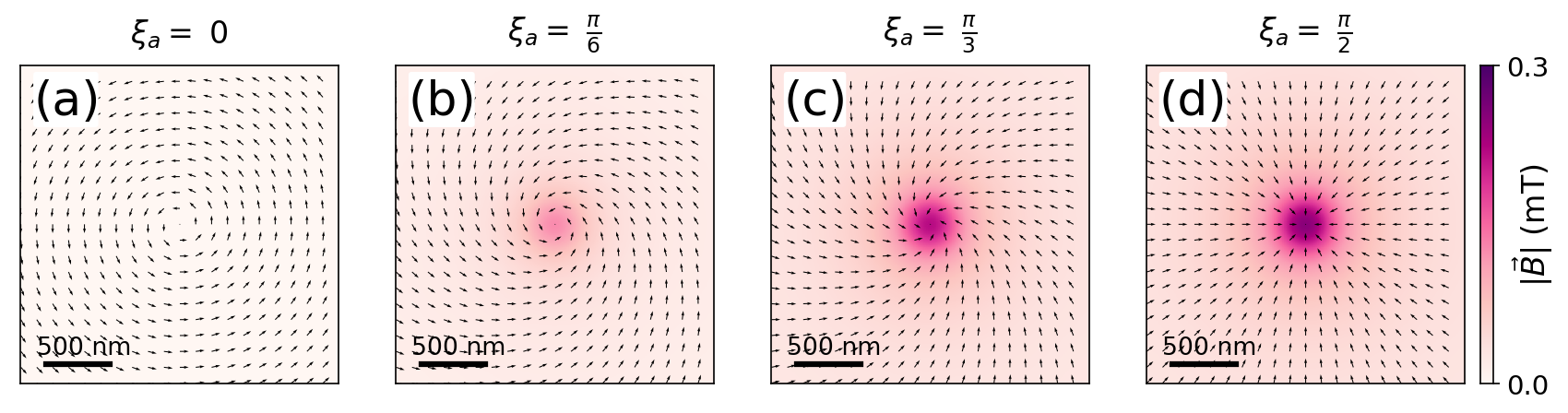}
    \caption{ \textbf{Magnetic field magnitude of merons with different phase $\xi_a$.} Field magnitude $|\vec{B}|$ obtained from the magnetic field simulations of merons with different $\xi_a$. Black arrows illustrate \mxy.}
    \label{fig:sim_b_mag}
\end{figure}

\begin{figure}
    \centering
    \includegraphics[width =  0.9\textwidth]{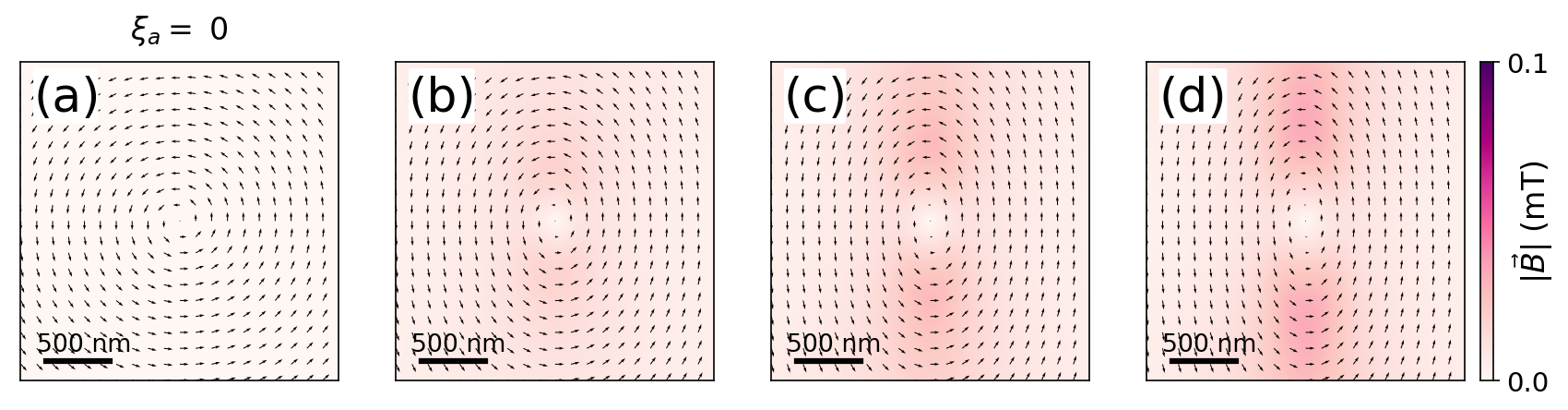}
    \caption{ \textbf{Distortion of a divergence-free texture.} Field magnitude $|\vec{B}|$ exhibited by an a-N\'eel meron with increasing distortion to the surrounding \mxy distribution from (a) to (d). Black arrows illustrate \mxy.}
    \label{fig:sim_b_mag_distort}
\end{figure}

For a divergence-free texture (i.e. $ik_x \tilde{m_x}+ik_y \tilde{m_y} = 0$), the z-component of the field in Eq. \ref{seq:bz_fourier} vanishes, rendering it invisible to magnetic field detection. This is further illustrated in Figs. \ref{fig:sim_b_mag_adw} and \ref{fig:sim_b_mag}, where ADWs and merons turn invisible when $\xi_a=0,\pi$. In reality, textures are never perfectly isolated, and the distortion produced by the proximity of other textures will make even a-N\'eel meron visible through the resulting magnetic field. This is illustrated in figure \ref{fig:sim_b_mag_distort}, which display the magnetic field of an a-N\'eel meron undergoing progressive geometrical distortion, and will be further discussed in Fig. \ref{fig:charge_analysis} in the context of a-N\'eel bimerons. 

\subsection{Magnetic Field and Charge of AFM Bimerons}
\label{si-sec:meron_class}
\begin{figure}
    \centering
    \includegraphics[width =  0.9\textwidth]{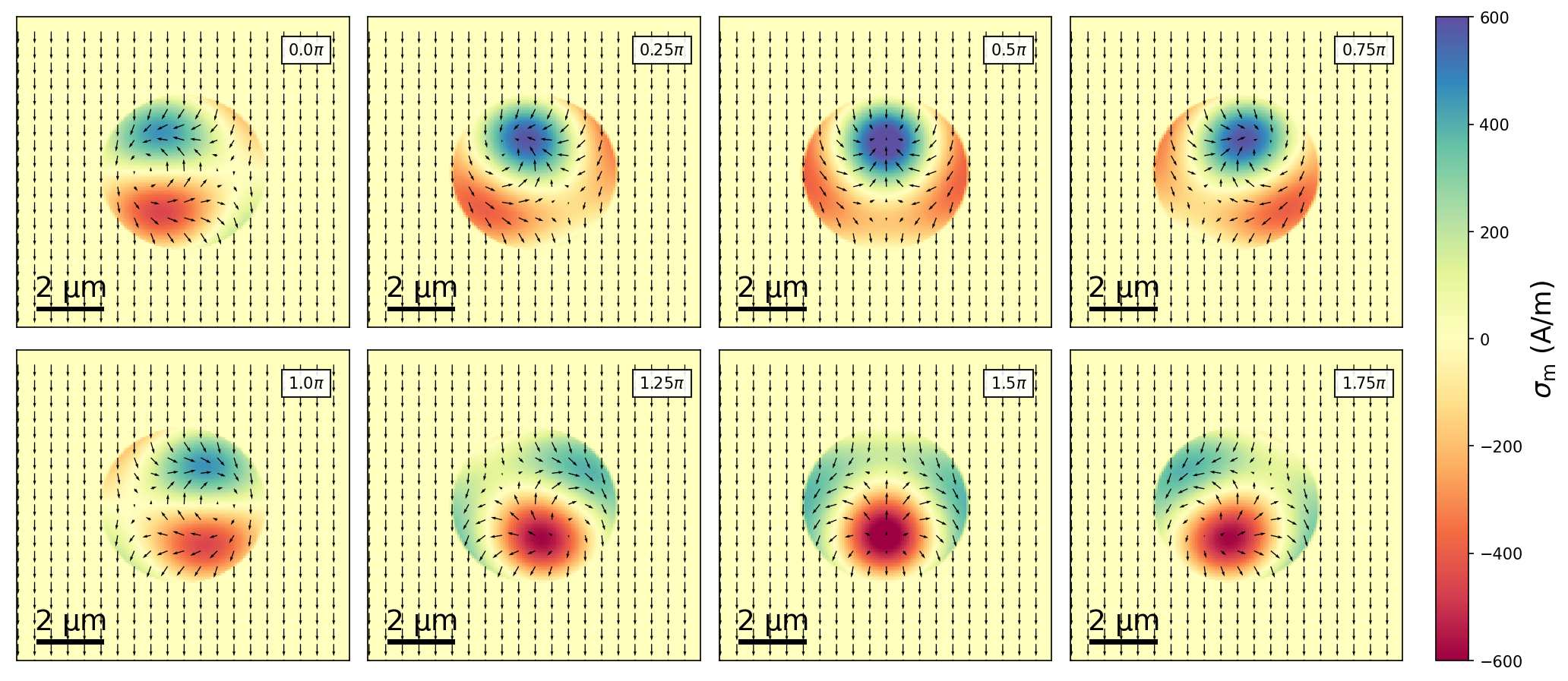}
    \caption{ \textbf{Magnetic charge density $\sigma_m$ of an AFM bimeron with varying phase $\xi_a$.} Black arrows illustrate \mxy.}
    \label{fig:bimeron_rho}
\end{figure}

An AFM bimeron in a uniform \mxy~background can be described analytically as a skyrmion model rotated by $90^\circ$ along an in-plane axis. Assuming a phase $\xi_a$ and radial size $R_B$, we have:
\begin{equation}
    \vec{l}  = 2~M_s [~\cos{\theta},~\sin{\theta} \sin(\phi+\xi_a),~-\sin{\theta} \cos(\phi+\xi_a)]~, \\
\label{seq:l_bimeron}
\end{equation}
\begin{equation}
    \vec{m}  = 2\Delta M_s ~[-~\sin{\theta} \sin(\phi+\xi_a),~\cos{\theta},~0~]~, \\
\label{seq:m_bimeron}
\end{equation}
\begin{align*}
    \mathrm{where,~} 
    \theta = \begin{cases} 
    \frac{\pi r}{R_B} &,~r\leq R_B\\
    \pi &,~r>R_B\\
    \end{cases}~.
\end{align*}
Hence,
\begin{align}
    \dmxy = \begin{cases}
    2\Delta M_s \big[-\frac{\pi}{R_B} \big( \cos{\theta}\cos{\phi}\sin{(\phi+\xi_a)+\sin{\theta}\sin{\phi}}\big) &~\\ ~~~~~~~~~~~~~~~~~~~~~~~~~~~~~~~~~~~~~~+\frac{1}{r}\sin{\theta}\sin{\phi}\cos{(\phi+\xi_a)}\big] &,~r\leq R_B\\
    0 &,~r>R_B\\
    \end{cases}~.
\label{seq:divm_bimeron}
\end{align}
Equation \ref{seq:divm_bimeron} allows us to model the $\sigma_m$ distribution of bimerons, see Fig. \ref{fig:bimeron_rho}. Here, we observe that individual monopolar and quadrupolar $\sigma_m$ distributions contributed by the meron and antimeron components, albeit significantly distorted in comparison to their isolated counterparts, can be discerned in the $\sigma_m$ distribution of an isolated AFM bimeron. This is slightly more evident in the case of a-Bloch bimerons ($\xi_a=\pi/2$, $3\pi/2$). For a-N\'eel bimerons the $\sigma_m$ distribution only reflects a strongly distorted quadrupolar character as the meron component is relatively divergence-free. 

\begin{figure}
    \centering
    \includegraphics[width =  0.9\textwidth]{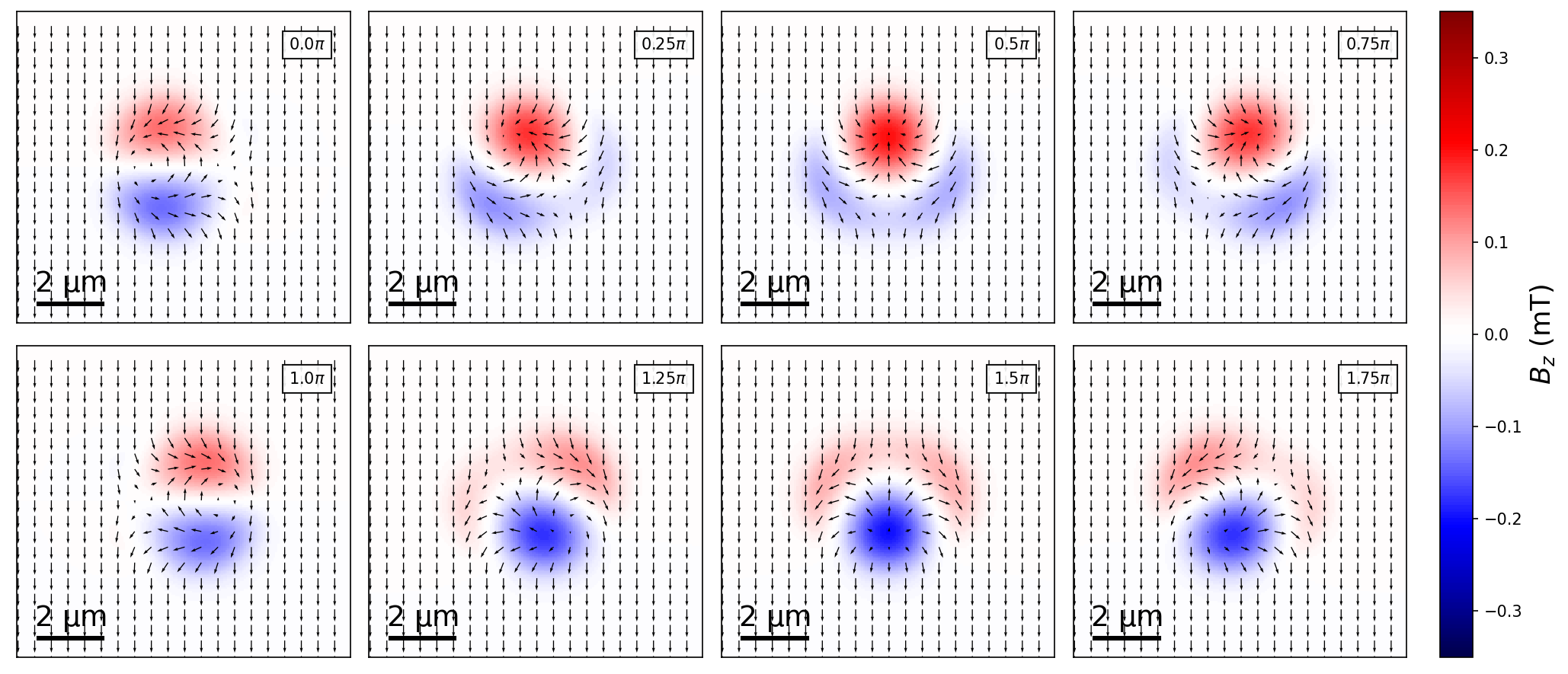}
    \caption{ \textbf{Magnetic field distribution of an AFM bimeron with varying phase $\xi_a$.} Black arrows illustrate \mxy.}
    \label{fig:bimeron_Bz}
\end{figure}

More importantly, we observe a non-zero $\sigma_m$ distribution, irrespective of the $\xi_a$ value, implying that bimerons of all characters generate distinct and detectable $B_z$ signatures, see Fig. \ref{fig:bimeron_Bz}. This enables DQM to play a versatile role in the future discovery and study of isolated AFM bimerons in engineered AFM systems with a defined in-plane anisotropy and inhomogeneous DMI. Interestingly, while the integrated charge $Q_m$ over a sufficiently large area centred at the core of the bimeron vanishes for all $\xi_a$, the asymmetric $\sigma_m$ distribution suggests that it may be possible to spatially manipulate AFM bimerons via external magnetic fields.

\subsection{Magnetic Field and Charge of AFM Skyrmions}
\label{si-sec:skyrmion_class}
\begin{figure}
    \centering
    \includegraphics[width =  0.9\textwidth]{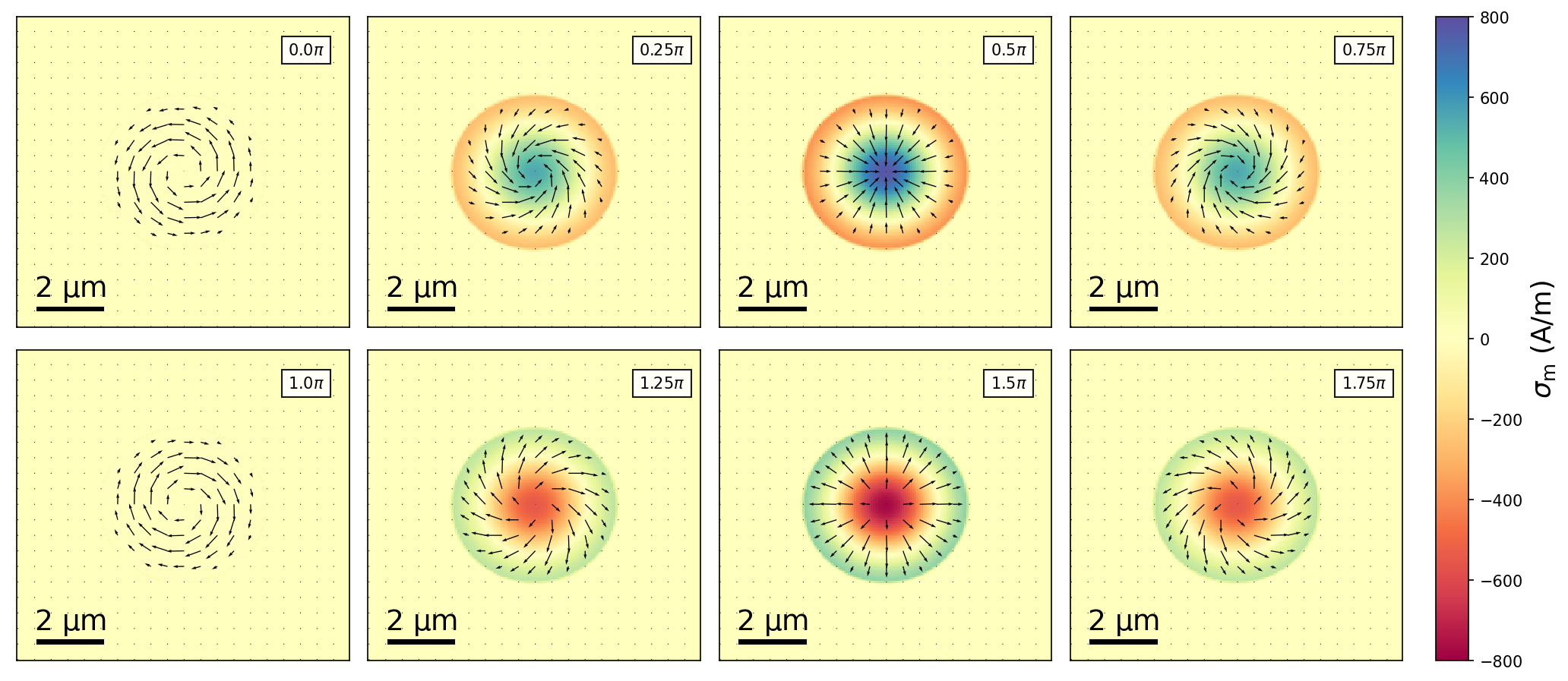}
    \caption{ \textbf{Magnetic charge density $\sigma_m$ distribution of an AFM skyrmion with varying phase $\xi_a$.} Black arrows illustrate \mxy.}
    \label{fig:skyrmion_rho}
\end{figure}

\begin{figure}
    \centering
    \includegraphics[width =  0.9\textwidth]{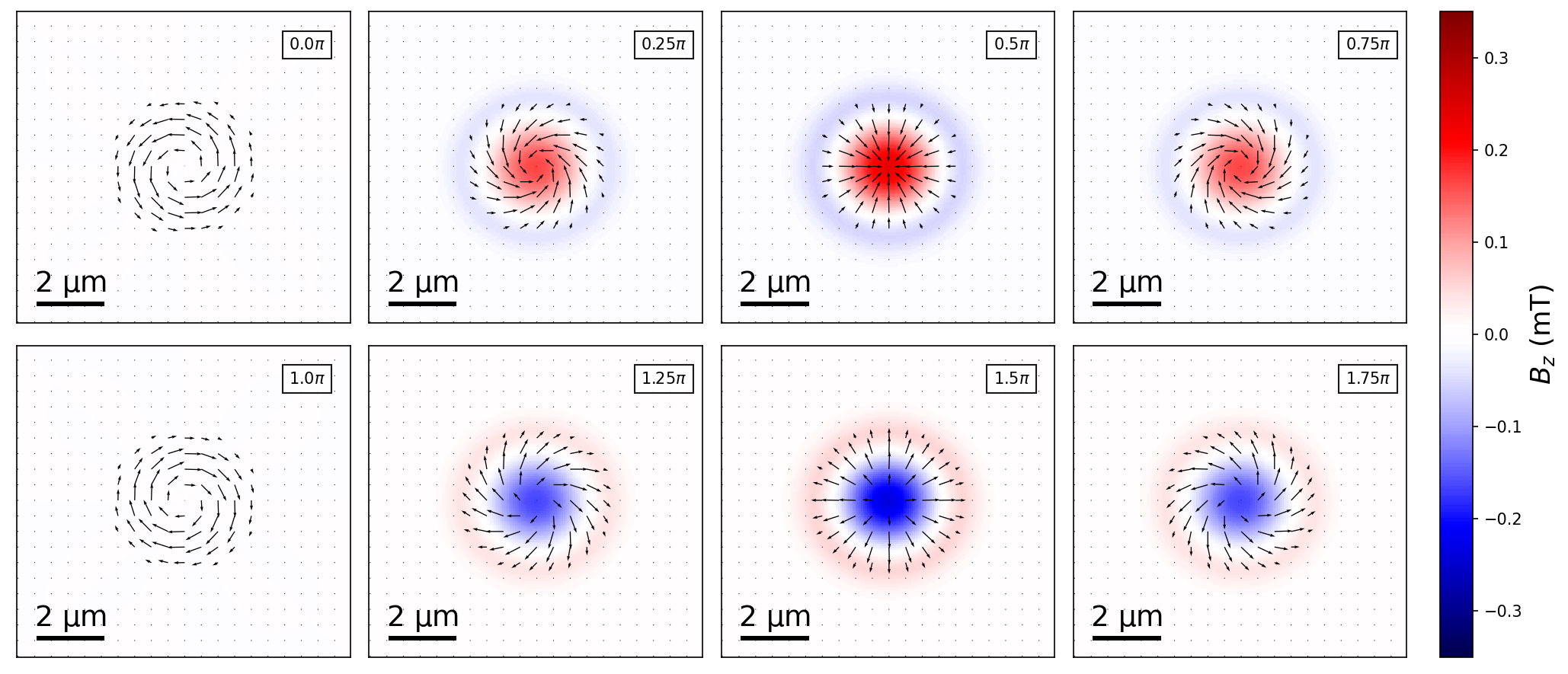}
    \caption{ \textbf{Magnetic field distribution of an AFM skyrmion with varying phase $\xi_a$.} Black arrows illustrate \mxy.}
    \label{fig:skyrmion_Bz}
\end{figure}

We can adapt the treatment from the previous section to construct the magnetic distribution in AFM skyrmions. Assuming a phase $\xi_a$ and radial size $R_S$, we have: 
\begin{equation}
    \vec{l}  = 2~M_s [~\sin{\theta} \cos(\phi+\xi_a),~\sin{\theta} \sin(\phi+\xi_a),~\cos{\theta}]~, \\
\label{seq:l_skyrmion}
\end{equation}\begin{equation}
    \vec{m}  = 2\Delta M_s [~-\sin{\theta} \sin(\phi+\xi_a),~\sin{\theta} \cos(\phi+\xi_a),~0]~, \\
\label{seq:m_skyrmion}
\end{equation}
\begin{align*}
    \mathrm{where,~} 
    \theta = \begin{cases} 
    \frac{\pi r}{R_S} &,~r\leq R_S\\
    \pi &,~r>R_S\\
    \end{cases}~.
\end{align*}
Hence,
\begin{equation}
    \dmxy = \begin{cases}
    2\Delta M_s \sin{\xi_a}\big( \frac{-\sin{\theta}}{r}- \frac{\pi\cos{\theta}}{R_S} \big) &,~r\leq R_S\\
    0&,~r>R_S\\
    \end{cases}~.
\end{equation}
AFM N\'eel skyrmions ($\xi_a = 0, \pi$) have vanishing $\sigma_m$, as they host divergence-free \mxy. Other skyrmions with $\xi_a\neq 0 , \pi$ have a non-zero $\sigma_m$ distribution, see Fig. \ref{fig:skyrmion_rho}. As expected, $B_z$ mirrors the $\sigma_m$ trend, Fig. \ref{fig:skyrmion_Bz}. While AFM N\'eel skyrmions are invisible, DQM can still play a crucial role in imaging AFM skyrmions with mixed or a-Bloch characters, with the added advantage of reading out the projected vorticity via the strength of $B_z$. In actuality, N\'eel skyrmions undergo deformation/distortion \cite{dist_skyrmion}, which could result in detectable magnetic field signatures, akin to what is expected for distorted a-N\'eel merons, as illustrated in Fig. \ref{fig:sim_b_mag_distort}.

\subsection{Integrated Magnetic Charge of AFM Merons and Antimerons}
\begin{figure}
    \centering
    \includegraphics[width =  0.6\textwidth]{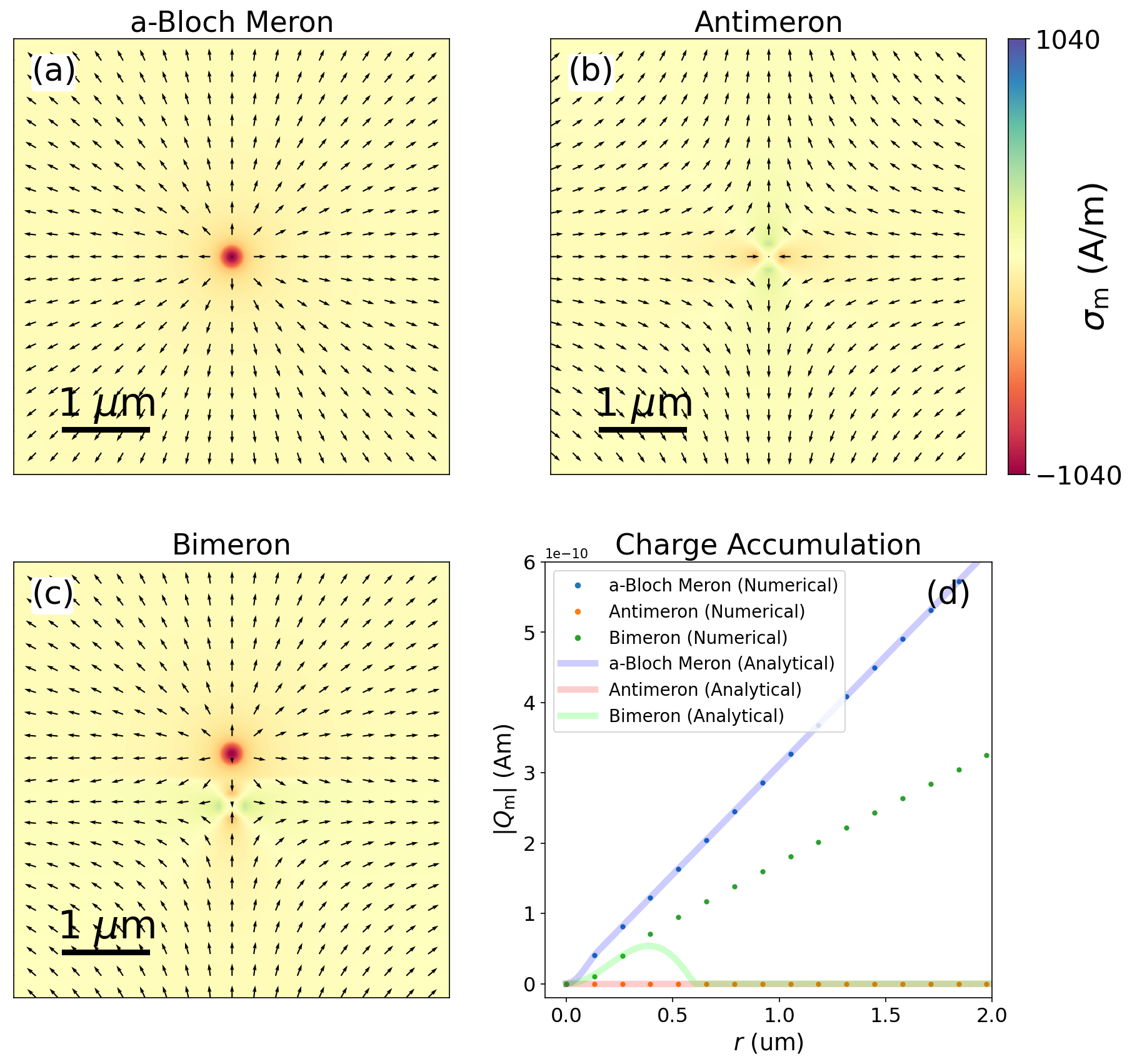}
    \caption{ \textbf{Charge accumulation of isolated textures.} Numerical simulation of charge density $\sigma_m$ distribution of (a) an a-Bloch meron, (b) an antimeron and (c) a bimeron. Arrows represent \mxy. (d) The total charge $|Q_m|$ accumulated in a circular area $S$ of increasing radius $r$, for each numerical simulation in (a-c), is given by the solid dots. $Q_m$ from analytical equations \ref{eq:total_charge} and \ref{eq:total_charge_bm} are represented by solid curves. The analytical $Q_m$ for a  a-Bloch meron and an antimeron ansatz agree well with the numerically obtained $Q_m$.}
    \label{fig:charge_analysis}
\end{figure}

The integrated charge $Q_m$ of an isolated (anti)meron with phase $\xi_a$ and winding number $\mathcal{N}$ as a function of radial distance $r$ from the core can be analytically written as:
\begin{align}
\label{eq:total_charge}
    Q_m (r) = 
    \begin{cases} 
    ~ 4\pi M_s~\Delta~\sin{(\xi_a)}~\sin{\big(\frac{\pi r}{2 R_M} \big)}~r~t &,~r\leq R_M,~\mathcal{N} = +1\\
    ~ 4\pi M_s~\Delta~\sin{(\xi_a)}~r~t&,~r> R_M,~\mathcal{N} = +1\\
    ~0~&, \mathcal{N} = -1
    \end{cases}~.
\end{align}
For isolated symmetric textures, $Q_m$ is either zero for antimerons ($\mathcal{N}=-1$) and a-N\'eel merons ($\mathcal{N}=+1$, $\xi_a=0,\pi$), or accumulates radially ($\mathcal{N}=+1, \xi_a\neq0,\pi$), as illustrated in Fig. \ref{fig:charge_analysis}(d). 

Moreover, the integrated charge $Q_m$ of an isolated bimeron with phase $\xi_a$ can be written as:
\begin{align}
\label{eq:total_charge_bm}
    Q_m (r) = 
    \begin{cases} 
    ~ 2\pi M_s~\Delta~\sin{(\xi_a)}~\sin{\big(\frac{\pi r}{R_B} \big)}~r~t &,~r\leq R_B\\
    ~ 0 &,~r> R_B
    \end{cases}~.
\end{align}
While isolated bimerons with $\xi_a\neq0,\pi$ accumulate some charge at short distances near their cores, see Fig. \ref{fig:charge_analysis}(d), the total integrated charge, outside $R_B$, for all isolated bimerons embedded in a uniformly aligned magnetic background is identically is zero.

\subsection{Charge Variability of AFM Bimerons}
\begin{figure}
    \centering
    \includegraphics[width =  0.9\textwidth]{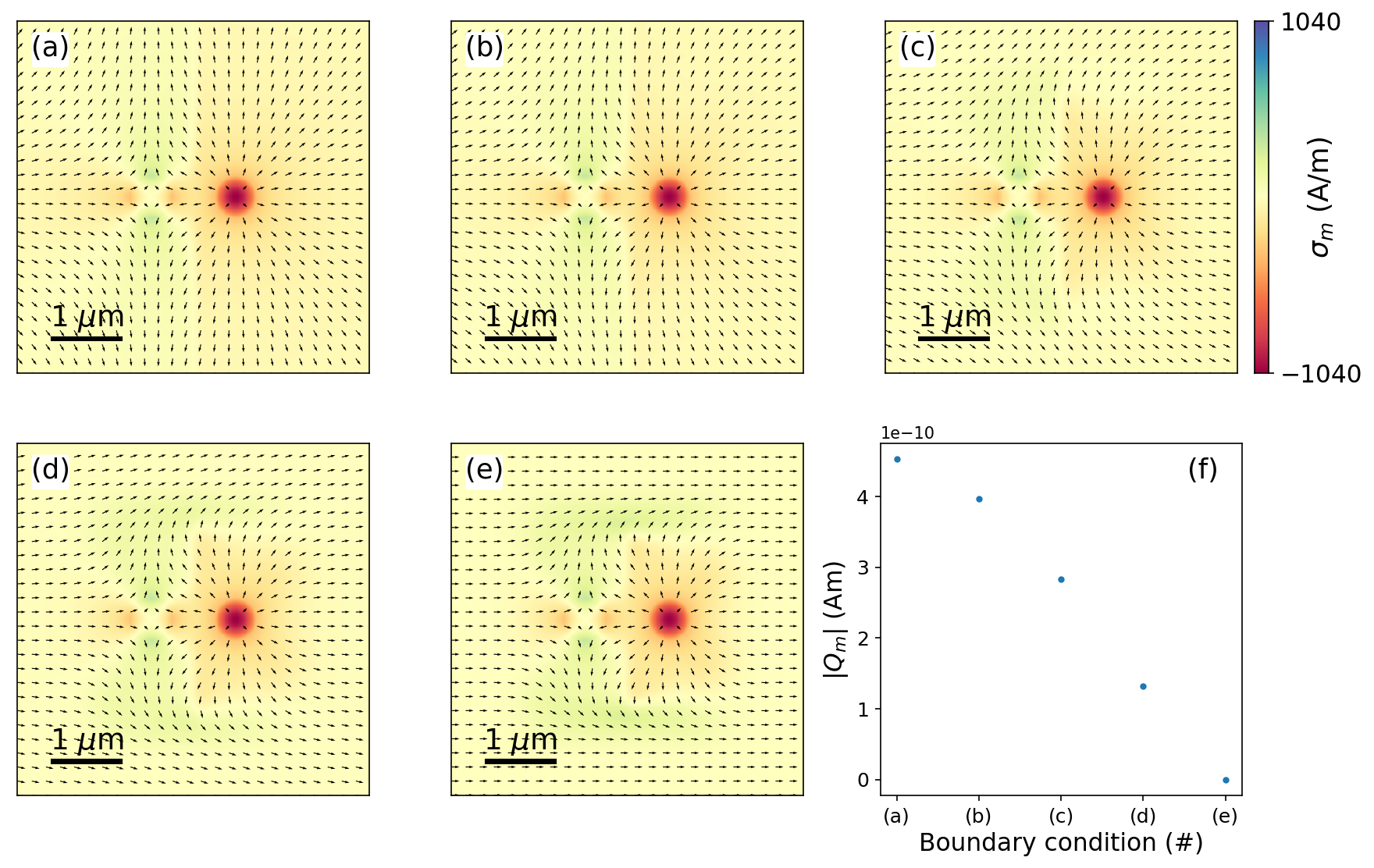}
    \caption{ \textbf{Charge variability of a bimeron.} (a-e) Numerical simulations of charge density $\sigma_m$ distribution of bimerons with various \mxy ~boundary conditions. Arrows represent \mxy. (f) Total charge $|Q_m|$ accumulated in a fixed area $S$ of various bimeron simulation with different \mxy~ boundary conditions in (a-e). For convenience, $S$ is defined as the square area covering the entirety of each panel, and is centered at the midpoint between the core of the meron and the antimeron.}
    \label{fig:charge_deform}
\end{figure}

The total charge $Q_m$ of a bimeron in a fixed area $S$ can be varied with different \mxy~ boundary conditions. Figs. \ref{fig:charge_deform}(a-e) illustrate bimerons with various boundary conditions, alongside their \mxy~and $\sigma_m$ distributions. Figure \ref{fig:charge_analysis}(f) shows the variability of $|Q_m|$ corresponding to panels (a-e), where each $|Q_m|$ is obtained from the integration of their respective $\sigma_m$ distribution over the whole region shown. Notably, in panel (e), the bimeron is embedded in a uniform magnetization with a boundary condition that satisfies:
\begin{equation}
    Q_m = \int_S \sigma_m ~ds 	\propto \oint_C \mxy \cdot \hat{e}~ dc = 0,
\end{equation}
where $C$ is the boundary of the area of integration $S$ and $\hat{e}$ is the unit normal to $C$.

\section{Meron Core Size Analysis}
\label{si-sec:size}
\begin{figure}
    \centering
    \includegraphics[width =  0.6\textwidth]{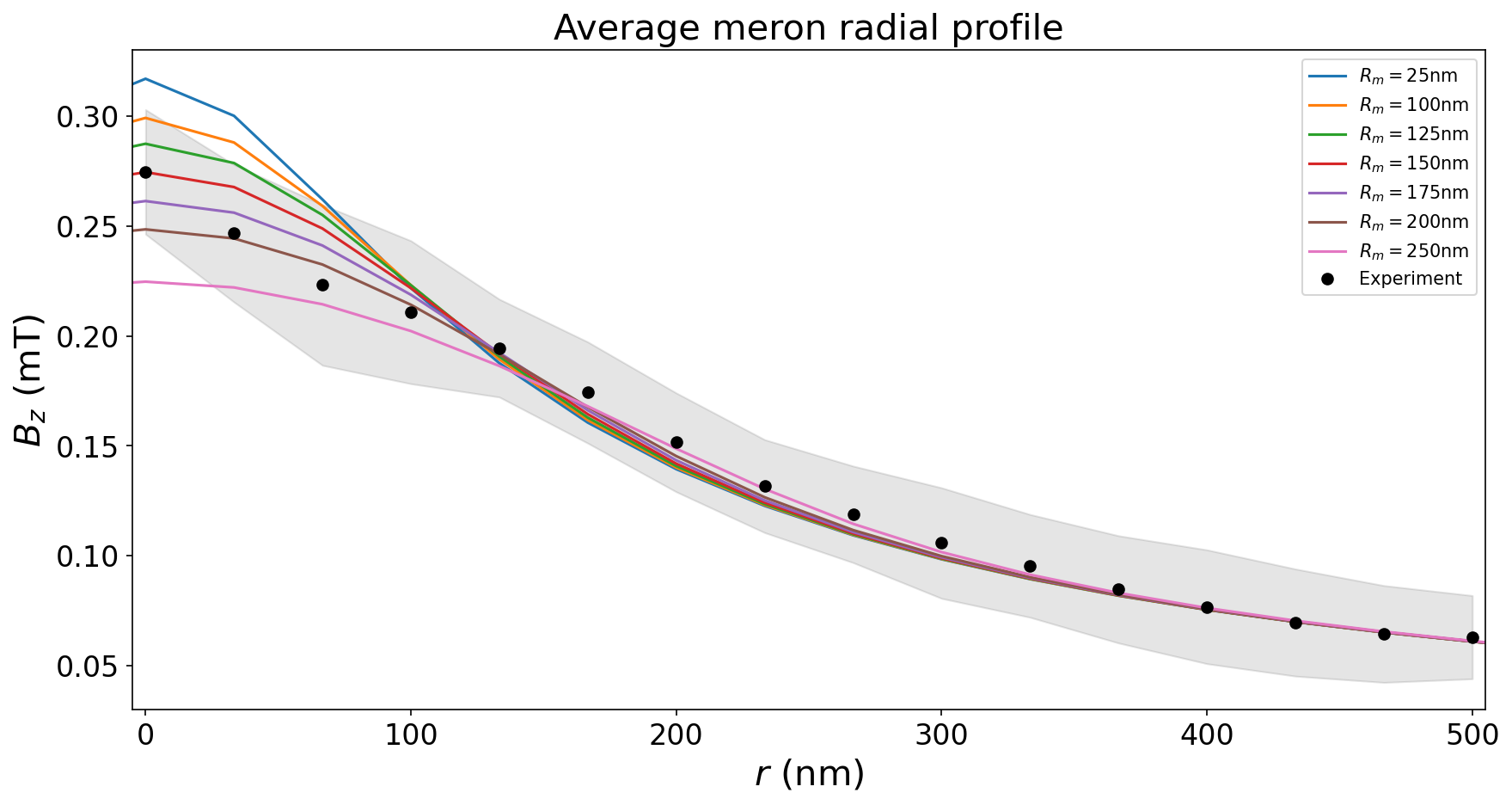}
    \caption{ \textbf{Average meron core size.} The average experimental $B_z$ profile of six merons as a function of radial distance $r$ from the center of the meron core (solid black dots, shaded area is the standard deviation). The experimental profile is compared with the simulated $B_z$ profiles of merons with various core sizes $R_M$ (coloured lines).}
    \label{fig:meron_core_size}
\end{figure}

The average meron radial profile shown in Fig. \ref{fig:meron_core_size} is obtained from the merons found in main text Fig. 1(h), by averaging their azimuthally--averaged radial $B_z$ profiles. In general, there is good qualitative agreement with the simulated $B_z$ profiles obtained from the structures in section \ref{si-sec:linear_model}. Any quantitative deviations are expected to result from the limitations of the linear meron ansatz and of a comparison between experimental data obtained from a composite textural fabric and an isolated meron model. We estimate $R_M \sim 150 \pm 40$ nm based on the maximum value of $B_z$, which corresponds reasonably well with our previous work \cite{Jani2021}.

\section{Retrieving \mxy}
\label{si-sec:m_reconstruction}
The inverse magnetostatic problem of retrieving $\vec{m}$ from $\vec{B}$ is in general ill posed. However, in our case, we can take advantage of existing constrains to approximate a solution. We first simplify the problem, as discussed in section \ref{si-sec:m_reconstruction_simple}, by utilizing the fact that the magnetic field components are linearly dependent, and that
the out-of-plane magnetization component vanishes in \feo. Next in section \ref{si-sec:m_reconstruction_reg}, we adopt appropriate regularisation strategies to obtain \mxy~ from $B_z$. Finally, some examples are discussed in sections \ref{si-sec:ADW_fit} and \ref{si-sec:meron_fit}. 

\subsection{Simplification of Retrieval Problem}
\label{si-sec:m_reconstruction_simple}
Retrieving $\vec{m}$ from $\vec{B}$ at distance $d$ above the sample surface (i.e., inversion of Eq.  \ref{seq:diptensor} ) is an ill-posed problem and is therefore not uniquely solvable. The problem can be simplified to one field component, since the other two are linearly dependent on the first one. We chose to work with $B_z$ since it intuitively reveals a low-pass filtered \dmxy~(eqn. \ref{seq:bz_real}). The inverse of the map $\boldsymbol{f}$ such that $B_z = \boldsymbol{f}\left(\vec{m}\right)$, is not still not unique, even within the restricted class of problems represented by the dipolar tensor (eqn. \ref{seq:diptensor}). However, a choice of appropriate constraints reduces the degrees of freedom, thereby allowing us to estimate $\vec{m}$ from a DQM scan. In the present case, we set the out-of-plane magnetization component $m_z = 0$. The inversion problem is thus reduced from three to two dimensions, but the reduced dimensionality is still not sufficient to estimate the in-plane magnetization vector. This is clear when considering equation \ref{seq:bz_fourier}, where we show that any divergence-free \mxy~, including trivial ordering, produces the same field distribution ($|B| = 0$, section \ref{ssec:div-zero}). 
\subsection{Regularization}
\label{si-sec:m_reconstruction_reg}
We adopt several regularisation strategies developed by the Lorentz transmission electron microscopy community to approximate a solution. Closely following the strategy in \cite{Caron2018}, we replace the ill-posed inversion problem with the following minimization problem:
\begin{equation}
    \min\left(\| \mathbf{F} \mathbf{m} - \mathbf{B}_z\| ^2  + \|\mathbf{\Gamma}\mathbf{m}\|^2 \right)~,
\label{seq:min}\end{equation}
where, $\mathbf{F}$, $\mathbf{B}_z$ and $\mathbf{m}$  are the matrix representation of the map $\boldsymbol{f}$, the observed $B_z$ and the reconstruction of $\vec{m}$, respectively. The spatial discretisation, implicit to the minimisation equation and also to a DQM scan, reduces the problem to finite dimensions and limits the maximum resolvable signal frequency to the inverse of the spatial resolution (i.e. $<\frac{1}{33}~\mathrm{GHz}$ in our images). The first term in equation \ref{seq:min} is the squared error needed for the base minimisation. The second term, known as the Tikhonov regularisation, is used to impose a smoothness constraint to the solution $\mathbf{m}$. This is achieved via $\mathbf{\Gamma} \mathbf{m}$ where $\mathbf{\Gamma}=\lambda\nabla$ is the matrix representation of the gradient $\nabla$ weighted by the constant $\lambda$. We constrain the problem further by fixing the magnitude of the canted magnetization $|\mxy|$. Given that $\mxy$ can be written as $|\mxy|(\cos{\phi_m},\sin{\phi_m})$, one only needs to reconstruct the angle $\phi_m$. Note that the reduction of the problem to one dimension alone does not guarantee a unique solution as we have a non-trivial null space. Finally, to quicken the minimisation and to avoid local minima, we initialize the problem with a $\mathbf{m}$ guess based on \textit{a priori} information of our system \cite{Jani2021}. $\phi_m$ is initialized based on the expected topological texture associated with the characteristic $B_z$ distribution shown analytically and via simulations in Fig. 2 in the main text, whereas $|\mxy|$ is fixed based on the linear meron ansatz (section \ref{si-sec:linear_model}) with an average core size of 150 nm (section \ref{si-sec:size}), and the linear ADW ansatz with a similar domain wall width (section \ref{seq:m_adw}). In this respect, we only attempt to reconstruct \mxy~for AFM textures and their immediate neighbourhood with characteristic $B_z$ signatures. Additionally, we minimize the problem with $B_z$ at $z = d_{NV}$ rather than $z = 0$ to avoid using low pass filters for the propagation, and with Neumann boundary condition to reduce edge effects.
\subsection{Fitting of Antiphase Domain Walls}
\label{si-sec:ADW_fit}
\begin{figure}
    \centering
    \includegraphics[width =  0.75\textwidth]{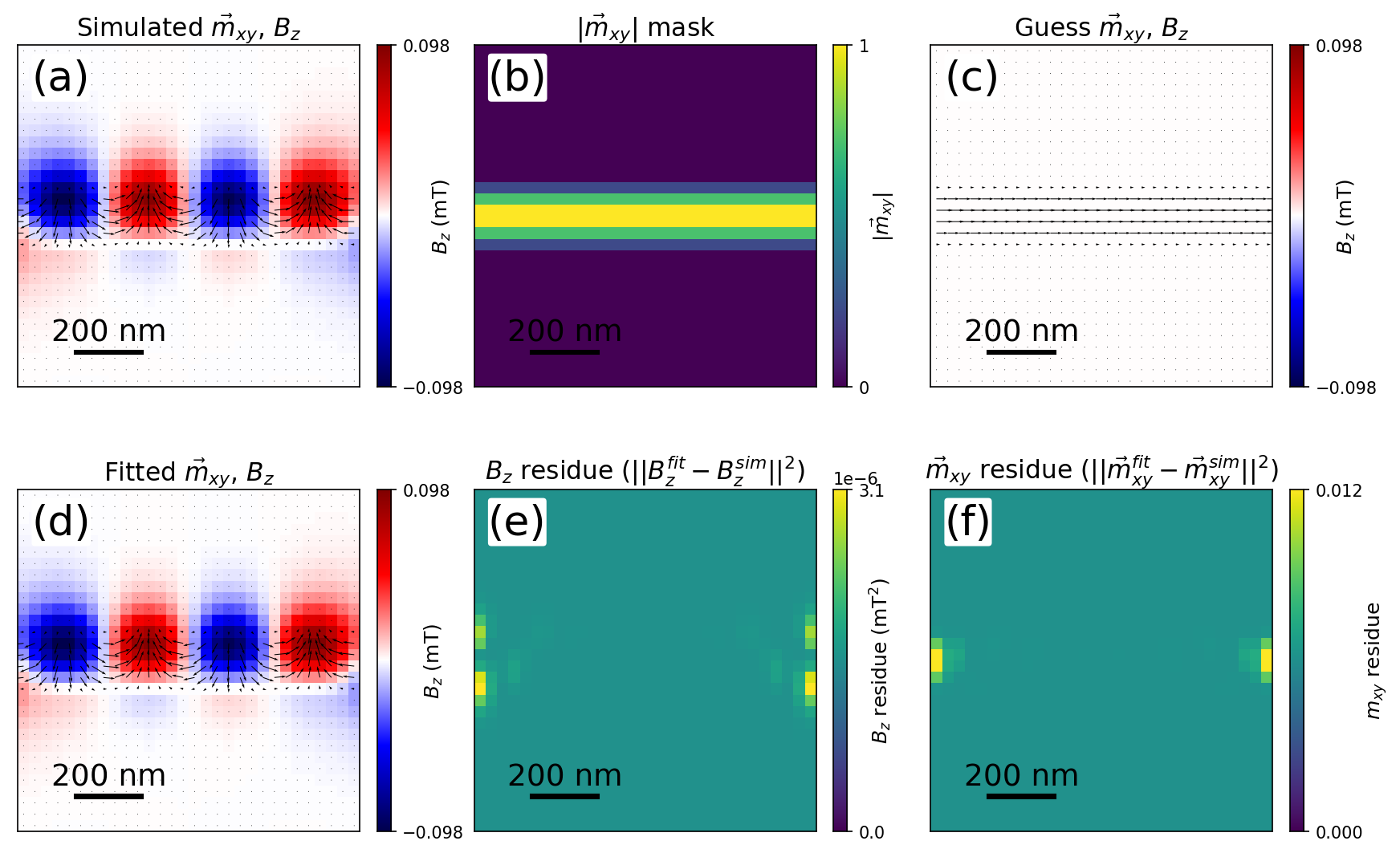}
    \caption{ \textbf{Simulated ADW fit.} \textbf{(a)} Simulated $\mxy$ of an AFM ADW and the resultant $B_z$ at $d_{NV} = 70$ nm. \textbf{(b-c)} (b) Fixed $|\mxy|$ mask and (c) guess $\mxy$~distribution used to initialize the minimisation problem. \textbf{(d-f)} (d) Minimized $\mxy$ solution and respective (e) $B_z$ and (f) $\mxy$ residues. Black arrows illustrate \mxy.}
    \label{fig:sim_ADW}
\end{figure}

\begin{figure}
    \centering
    \includegraphics[width =  0.75\textwidth]{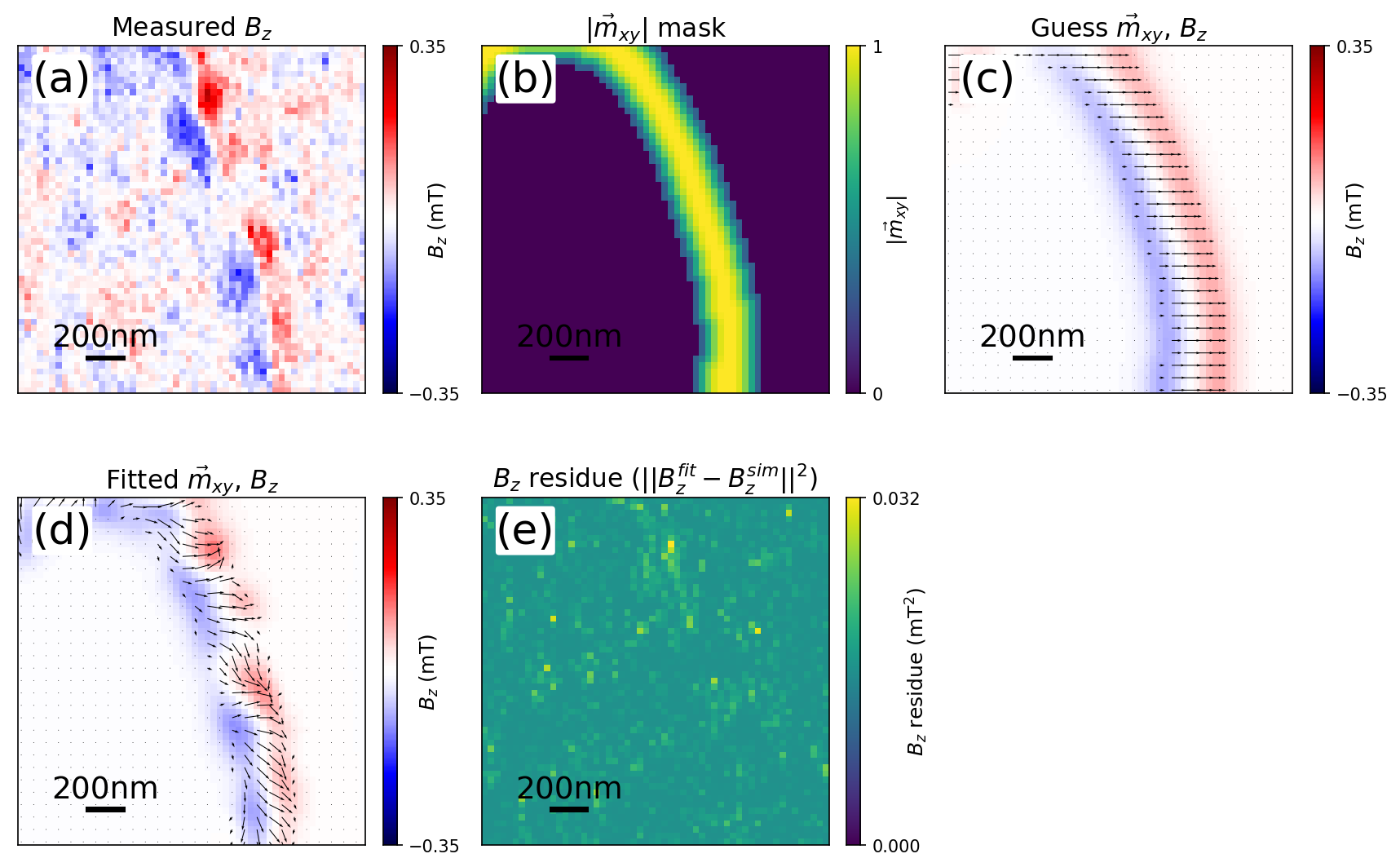}
    \caption{ \textbf{Experimental ADW fit.} \textbf{(a)} Measured $B_z$ of an AFM ADW at $d_{NV} = 70$ nm. \textbf{(b-c)} (b) Fixed $|\mxy|$ mask and (c) guess $\mxy$~distribution used to initialize the minimisation problem. \textbf{(d-f)} (d) Minimized $\mxy$ solution and respective (e) $B_z$ residue. Black arrows illustrate \mxy.}
    \label{fig:expt_ADW}
\end{figure}

We assume $|\mxy|= 0$  outside the ADW. Inside the ADW, $|\mxy|$ varies according to the linear ADW model (section \ref{seq:m_adw}). It is easy to separate these areas since out-of-plane AFM domains have negligible divergence while the ADWs have a distinct $B_z$ signature. We show an example of the reconstruction process from the $B_z$ image of a simulated multi-chiral ADW $\mxy$ distribution in Fig. \ref{fig:sim_ADW}(a). The $|\mxy|$ mask is obtained by estimating the binary mask from the simulated $B_z$, and then modulating it as per the linear ansatz, see Fig. \ref{fig:sim_ADW}(b). Here we initialize the problem with a trivial $\phi_m(x,y) = 0$ and the corresponding $\mxy$ guess is given in Fig. \ref{fig:sim_ADW}(c). The minimized $\mxy$ solution given in Fig. \ref{fig:sim_ADW}(d) faithfully reproduces the simulated $\mxy$ distribution, with errors localized at the edges as indicated by the fit residues in Figs. \ref{fig:sim_ADW}(e) and (f). Next, we apply the same protocol to the experimentally obtained $B_z$, with its respective $w\sim 2R_m$ and minimized $\mxy$ solution given in Figs. \ref{fig:expt_ADW}.

\subsection{Fitting of (Anti)Merons and Bimerons}
\label{si-sec:meron_fit}
\begin{figure}
    \centering
    \includegraphics[width =  0.75\textwidth]{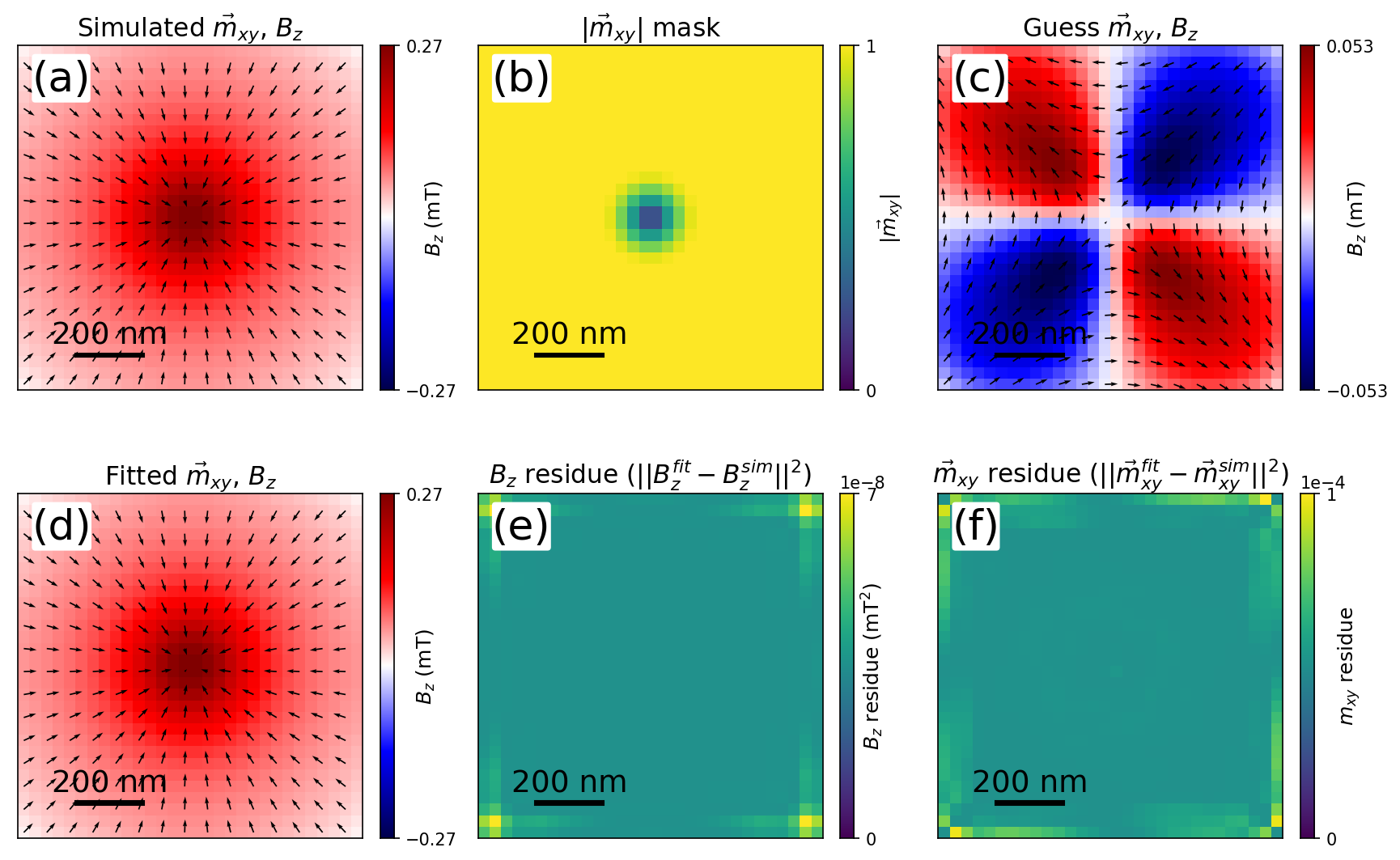}
    \caption{ \textbf{Simulated a-Bloch meron fit.} Simulated $\mxy$ of an AFM a-Bloch meron and the resultant $B_z$ at $d_{NV} = 70$ nm. \textbf{(b-c)} (b) Fixed $|\mxy|$ mask and (c) guess $\mxy$~distribution used to initialize the minimisation problem. \textbf{(d-f)} (d) Minimized $\mxy$ solution and respective (e) $B_z$ and (f) $\mxy$ residues. Black arrows illustrate \mxy.}
    \label{fig:sim_AM_to_M}
\end{figure}

\begin{figure}
    \centering
    \includegraphics[width =  0.75\textwidth]{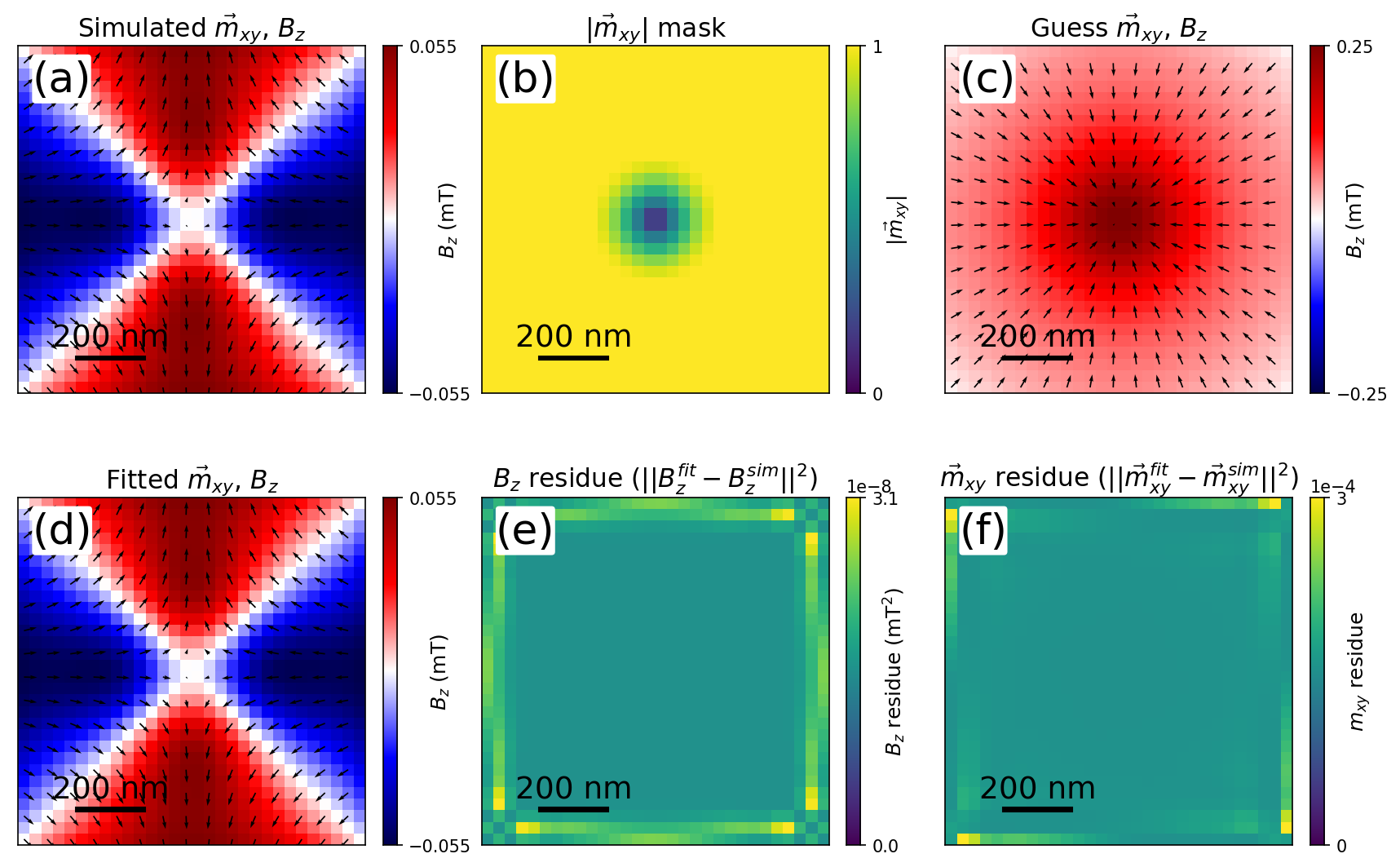}
    \caption{ \textbf{Simulated antimeron fit.} Simulated $\mxy$ of an AFM antimeron and the resultant $B_z$ at $d_{NV} = 70$ nm. \textbf{(b-c)} (b) Fixed $|\mxy|$ mask and (c) guess $\mxy$~distribution used to initialize the minimisation problem. \textbf{(d-f)} (d) Minimized $\mxy$ solution and respective (e) $B_z$ and (f) $\mxy$ residues. Black arrows illustrate \mxy.}
    \label{fig:sim_M_to_AM}
\end{figure}

Similar to the procedure in section \ref{si-sec:ADW_fit}, $|\mxy| = 1$ is assumed everywhere except in the (anti)meron core, which is modelled with a linear ansatz with $R_M = 150$ nm (section \ref{si-sec:size}). We show the \mxy~reconstruction of a simulated a-Bloch AFM meron and an AFM antimeron in Figs. \ref{fig:sim_AM_to_M} and \ref{fig:sim_M_to_AM}, respectively. In both cases, we start off with $\xi_a$ that deviates from the simulated texture, and the corresponding $\mxy$ guess is given in panel (c). The resultant minimized solution given in panel (d) closely reproduces the simulated $\mxy$ distribution in panel (a). The respective residues are given in panels (e-f).

\begin{figure}
    \centering
    \includegraphics[width =  0.75\textwidth]{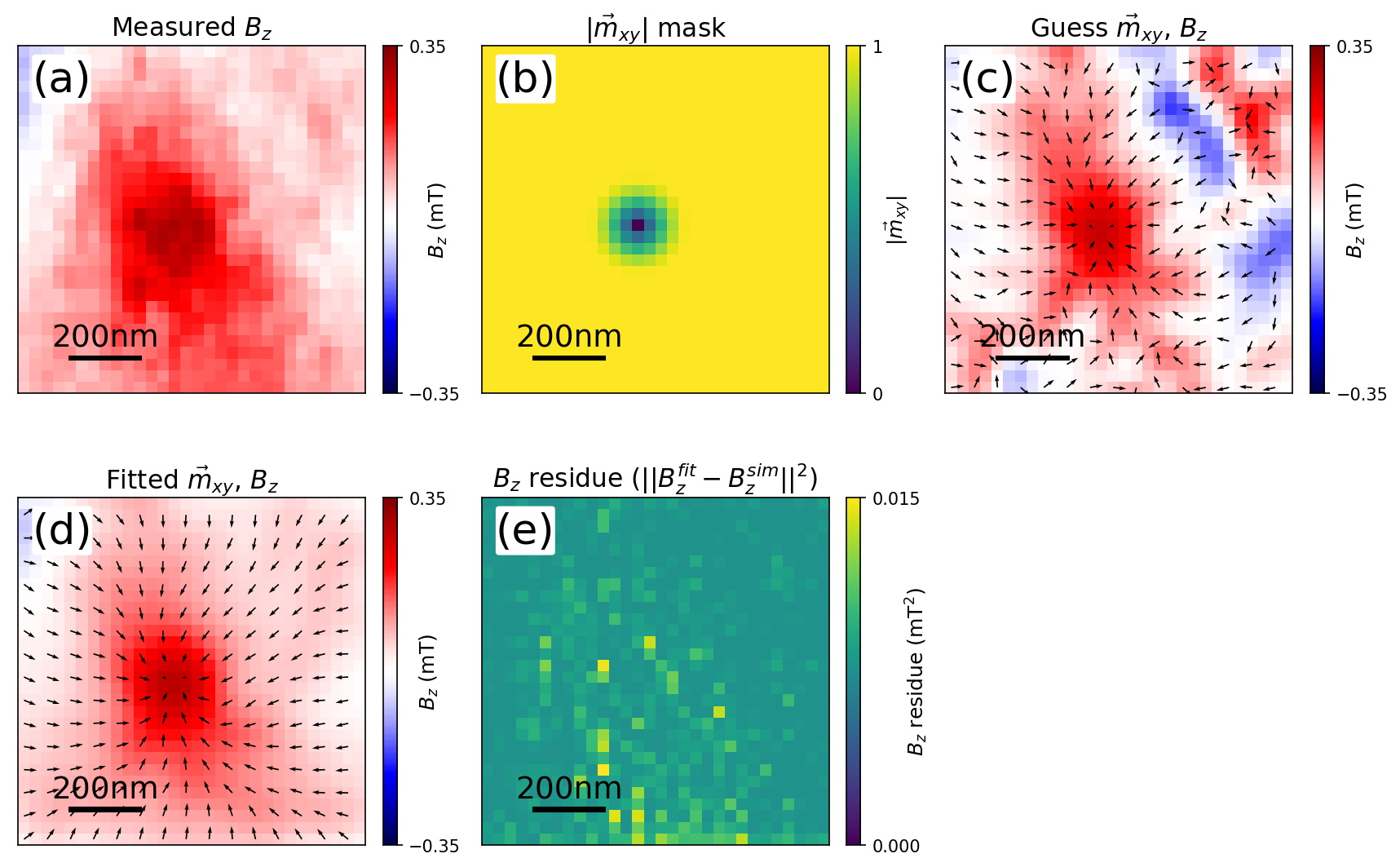}
    \caption{ \textbf{Experimental a-Bloch meron fit.}  Measured $B_z$ of an AFM right a-Bloch meron at $d_{NV} = 70$ nm. \textbf{(b-c)} (b) Fixed $|\mxy|$ mask and (c) guess $\mxy$~distribution used to initialize the minimisation problem. \textbf{(d-e)} (d) Minimized $\mxy$ solution and respective (e) $B_z$ residue. Black arrows illustrate \mxy.}
    \label{fig:exp_rm}
\end{figure}

\begin{figure}
    \centering
    \includegraphics[width =  0.75\textwidth]{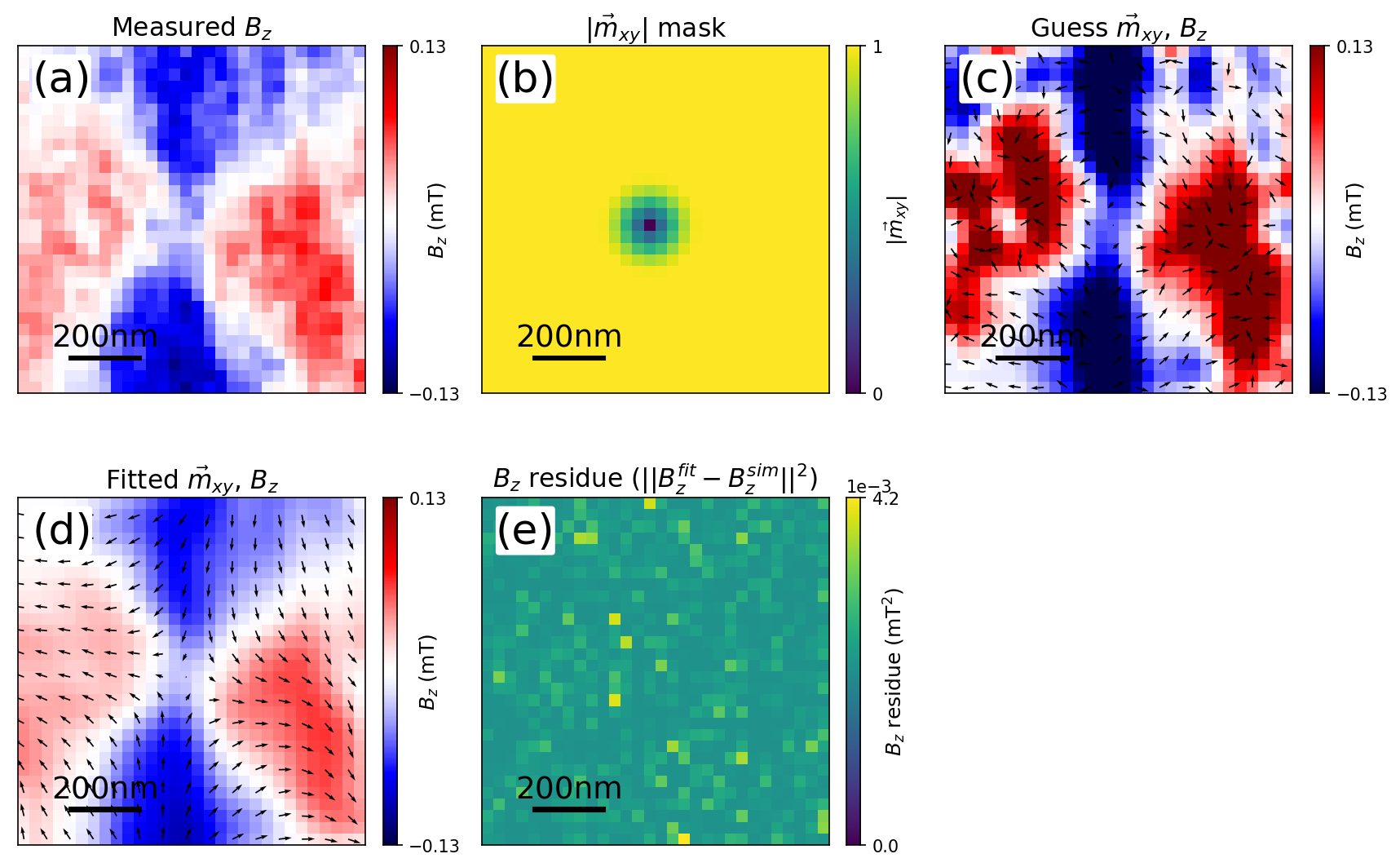}
    \caption{ \textbf{Experimental antimeron fit.}  Measured $B_z$ of an AFM antimeron at $d_{NV} = 70$ nm. \textbf{(b-c)} (b) Fixed $|\mxy|$ mask and (c) guess $\mxy$~distribution used to initialize the minimisation problem. \textbf{(d-e)} (d) Minimized $\mxy$ solution and respective (e) $B_z$ residue. Black arrows illustrate \mxy.}
    \label{fig:exp_am}
\end{figure}

\begin{figure}
    \centering
    \includegraphics[width =  0.75\textwidth]{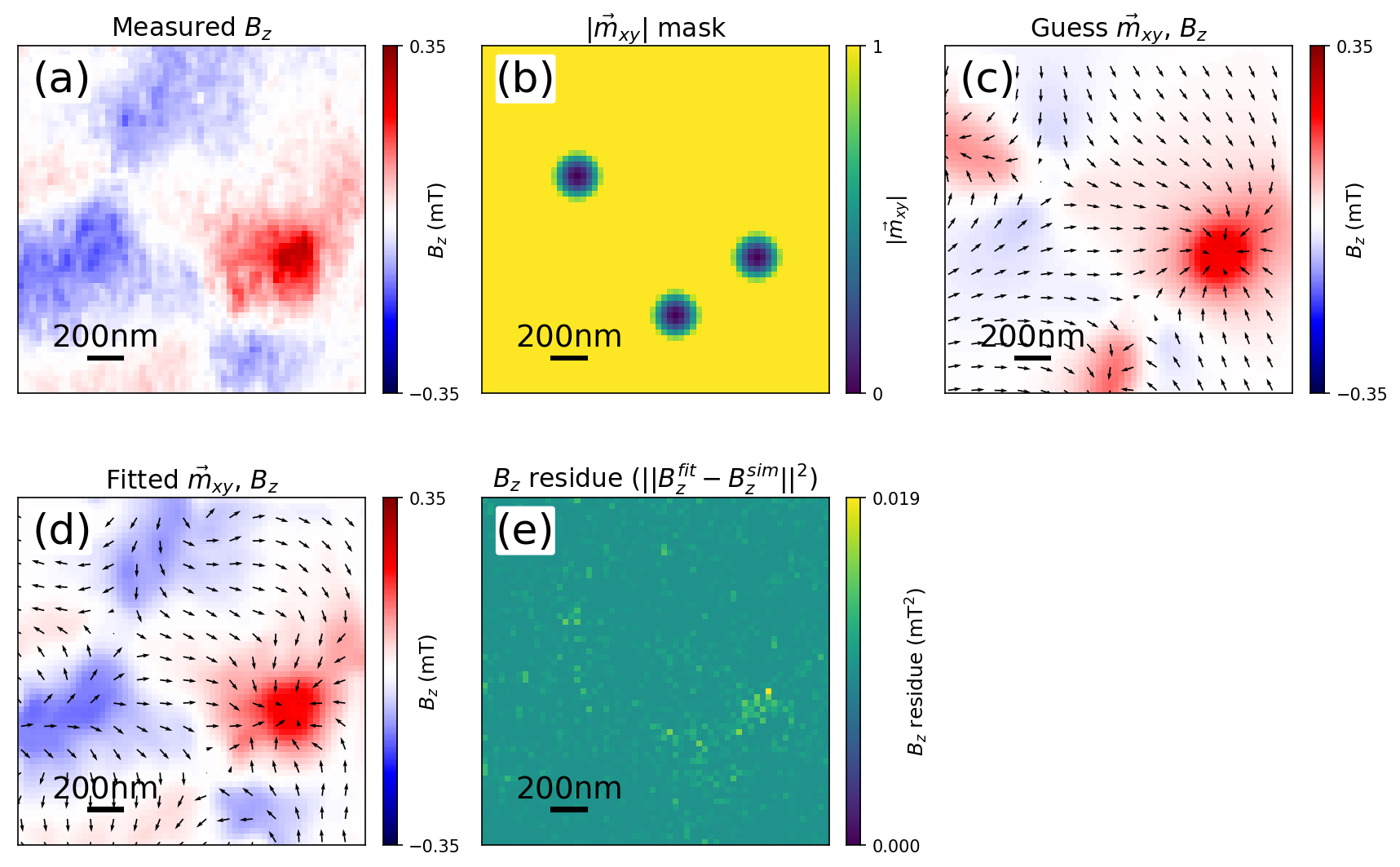}
    \caption{ \textbf{Experimental fit for composite texture.} Measured $B_z$ of a composite texture consisting of two antimerons and an a-Bloch meron at $d_{NV} = 70$ nm. \textbf{(b-c)} (b) Fixed $|\mxy|$ mask and (c) guess $\mxy$~distribution used to initialize the minimisation problem. \textbf{(d-e)} (d) Minimized $\mxy$ solution and respective (e) $B_z$ residue. Black arrows illustrate \mxy.}
    \label{fig:exp_bim}
\end{figure}

Finally, we apply the same reconstruction protocol to experimental images, from which we obtain the $\mxy$ fits of an anti-clockwise a-Bloch meron (Fig. \ref{fig:exp_rm}), an antimeron (Figure \ref{fig:exp_am}) and a composite structure consisting of one meron and two antimerons (Fig. \ref{fig:exp_bim}).
\subsection{Field imaging and Magnetisation Reconstruction of \feo}
Unique magnetisation reconstruction is a general problem in the stray-field imaging. Nonetheless, the rigorous identification procedure (see \ref{si-sec:distinguish}), together with the implementation of systematic regularization protocols, has helped constrain the magnetic solution space and provided meaningful canted moment reconstructions. While these reconstructions may not always be unique, they are in practice entirely consistent with AFM textures reported via X-ray imaging\cite{Jani2021}. 

Although X-ray linearly dichroic imaging is not affected by the same uniqueness problem, it suffers from an intrinsic $180 ^\circ$ ambiguity which would make reconstructions of the canted moment more ambiguous than in the present case. Additionally, the canted magnetisation in \feo~is too weak to produce circular dichroic X-ray contrast\cite{Jani2021}. Therefore, field imaging via DQM provides an invaluable view of the rich magnetic charge phenomenology in \feo.
\section{Reversal Field Imaging}
\label{si-sec:field_imaging}
\begin{figure}
    \centering
    \includegraphics[width =  0.9\textwidth]{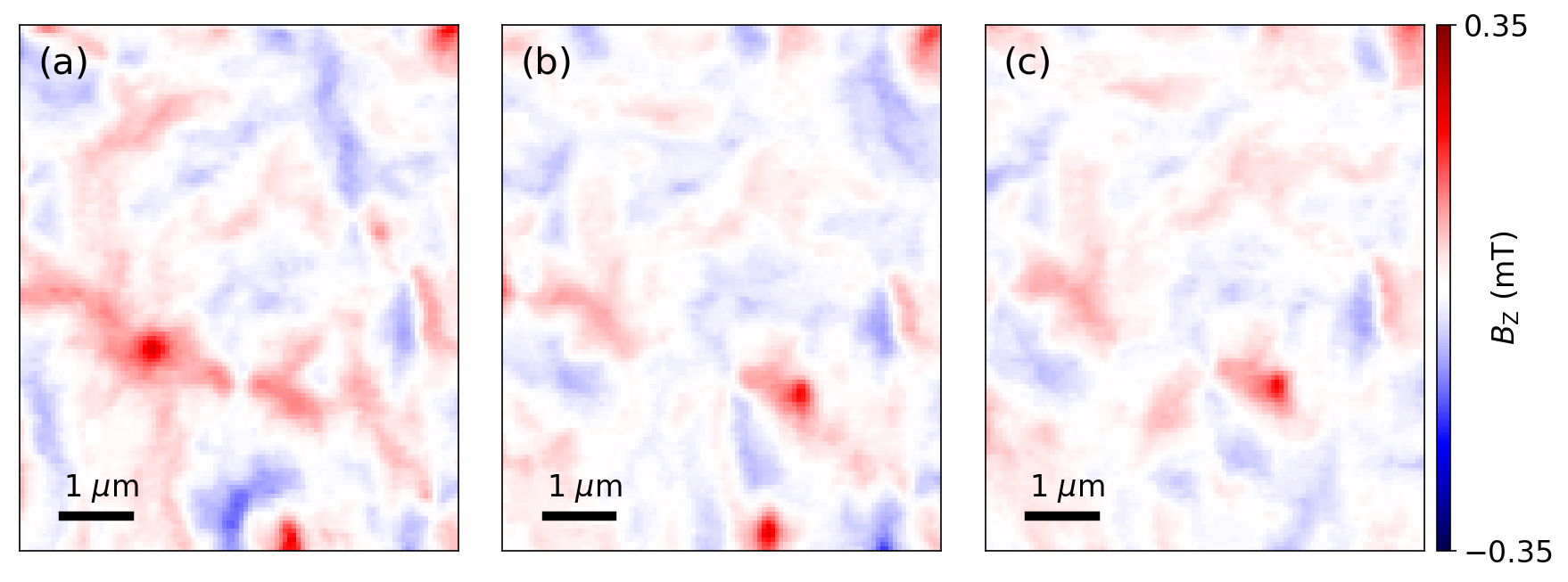}
    \caption{ \textbf{Reversal field imaging at remanence.} Remanent state of \feo~after applying an \textit{in-situ} in-plane magnetic field with magnitude varying from (a) 150mT, (b) 200mT and (c) 250mT.}
    \label{fig:field_dep}
\end{figure}
\begin{figure}
    \centering
    \includegraphics[width =  0.4\textwidth]{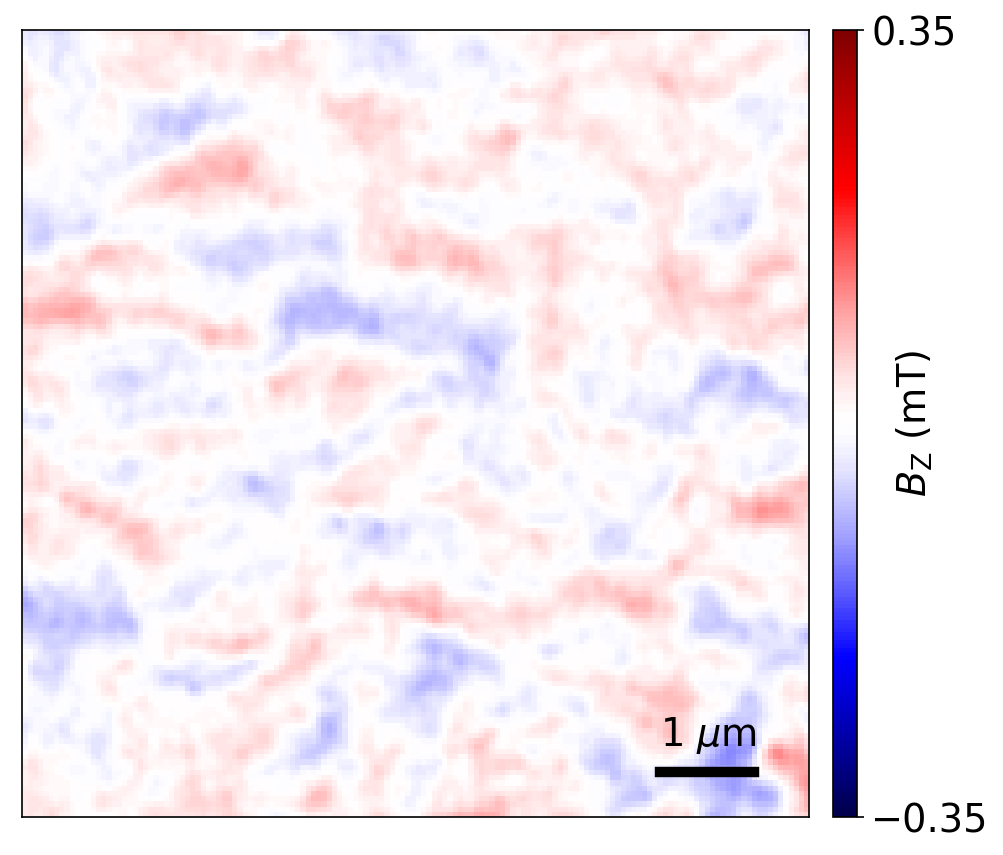}
    \caption{ \textbf{Annihilation of topological spin entities.} Remanent state of \feo~after applying an \textit{ex-situ} in-plane magnetic field of $\sim 400~ \mathrm{mT}$.}
    \label{fig:field_annihilation}
\end{figure}
Upon application of strong in-plane magnetic fields, it is possible to drive AFM domain realignment which results in meron-antimeron annihilation \cite{Jani2021}. To confirm this, two permanent magnet poles were aligned parallel to the sample surface resulting in a homogeneous field across the sample, variable via micrometer stages. Since imaging under high off-NV-axis field is not possible, the external field was applied for a few minutes followed by NV imaging in zero field, see Fig. \ref{fig:field_dep}. Although the scans are approximately in the same area, open-loop scanning requires post processing, wherein the images have been registered using distinct field features in the scans. Several merons and antimerons are initially present. As the magnitude of the external field increases, annihilation processes occur, accompanied by an alteration of the surrounding trivial AFM textures. Due to the limited field magnitude that can be applied \textit{in-situ}, large-scale annihilation of textures was driven by an \textit{ex-situ} 400 mT field \cite{Jani2021}, see Fig. \ref{fig:field_annihilation}.
Due to the probabilistic nature of the generation of topological textures via the Kibble-Zurek mechanism, the occurrence of merons with different projected vorticities (red vs blue cores) can vary across the sample \cite{Chmiel2018}. It is noted that while Fig. \ref{fig:field_annihilation} shows predominately anti-clockwise a-Bloch merons (red cores), additional large area scans (see S9) confirm approximately similar distributions of merons with different projected vorticities (red vs blue cores) are obtained due to the absence of chiral interactions in \feo \cite{Jani2021}.
\section{Additional Data}
\begin{figure}[H]
    \centering
    \includegraphics[width =  0.75\textwidth]{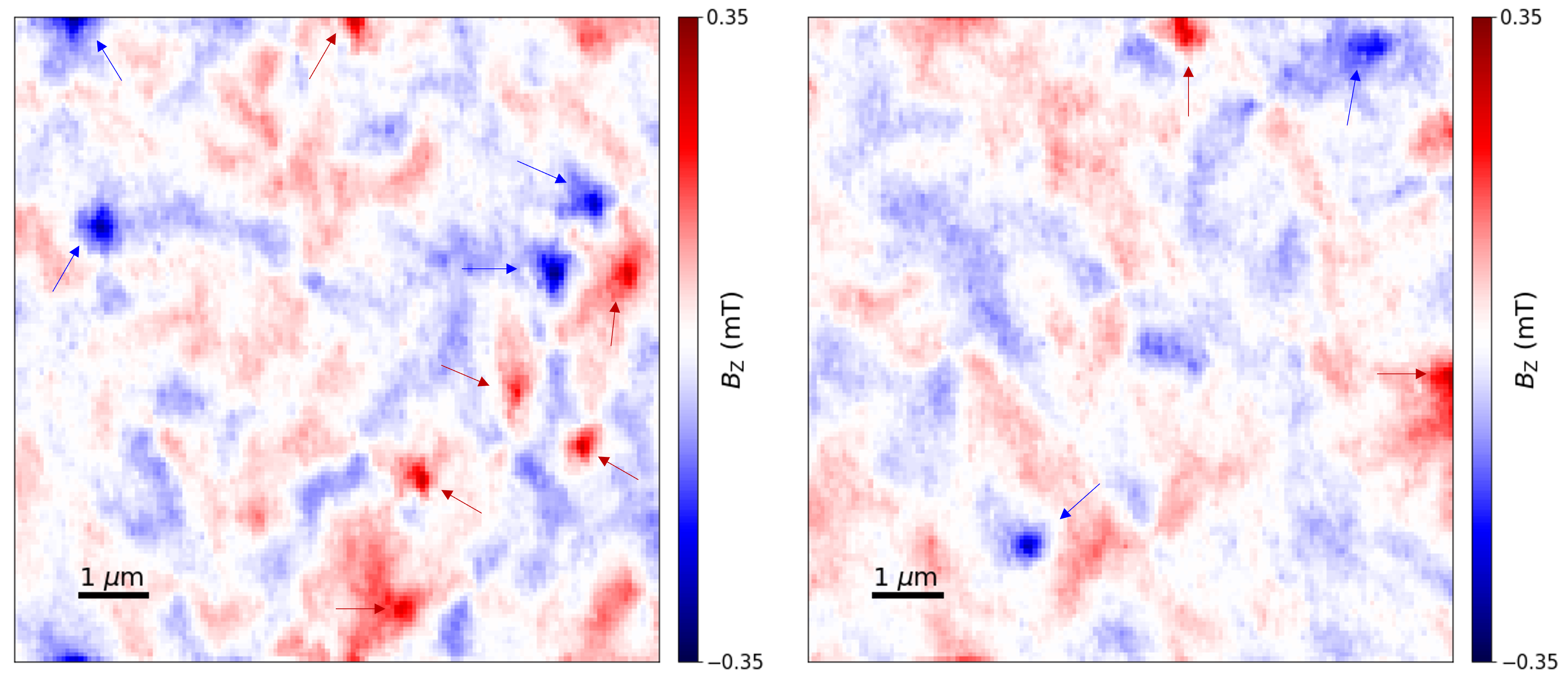}
    \caption{ \textbf{Large Area RT Images Taken After Thermal Cycling.} Field imaging over a large area showing presence of both right and left a-Bloch merons indicated by blue and red arrows respectively.}
    \label{fig:large_area}
\end{figure}

\end{document}